\documentclass{article}
\usepackage{amsmath}
\usepackage{amssymb}

\newtheorem{theorem}{Theorem}[section]
\newtheorem{lemma}{Lemma}[section]
\newtheorem{corollary}{Corollary}[section]
\newtheorem{prop}{Proposition}[section]

\newtheorem{definition}{Definition}[section]

\numberwithin{equation}{section}

\begin{document} 

\centerline{Direct and reverse log-Sobolev inequalities in} 
\centerline{$\mu$-deformed Segal-Bargmann analysis}

\vskip .5cm

\centerline{Carlos Ernesto Angulo Aguila \footnote{Research partially supported by CONACYT (Mexico) project 49187-F.}
} 

\centerline{Universidad Aut\'onoma de San Luis Potos\'{\i}}

\centerline{San Luis Potos\'{\i}, Mexico}

\centerline{email: anguloa@fciencias.uaslp.mx} 

\vskip .5cm

\centerline{Stephen Bruce Sontz\footnote{Research partially supported by CONACYT (Mexico) projects P-42227-F and 49187-F.}}

\centerline{Centro de Investigaci\'on en Matem\'aticas, A.C. (CIMAT)}

\centerline{Guanajuato, Mexico}

\centerline{email: sontz@cimat.mx}

\vskip .4cm

\centerline{Dedicated to the memory of Marvin Rosenblum}

\vskip .8cm

\begin{abstract}
\noindent 
Both direct and reverse log-Sobolev inequalities, relating the Shannon entropy
with a $\mu$-deformed energy, are shown to
hold in a family of $\mu$-deformed Segal-Bargmann spaces. 
This shows that the $\mu$-deformed energy
of a state is finite if and only if its Shannon entropy is finite. 
The direct inequality is
a new result, while the reverse inequality has already been shown by the authors
but using different methods.  
Next the $\mu$-deformed energy of a state is shown to be finite
if and only if its Dirichlet form energy is finite. 
This leads to both direct and reverse log-Sobolev inequalities that 
relate the Shannon entropy with the Dirichlet energy. 
We obtain that the Dirichlet energy
of a state is finite if and only if its Shannon entropy is finite. 
The main method used here
is based on a study of the reproducing kernel function of these spaces and 
the associated integral kernel transform. 
\end{abstract} 

\vskip .3cm \noindent
\textit{Keywords:} Segal-Bargmann analysis, log-Sobolev inequality, reverse log-Sobolev
inequality, reproducing kernel Hilbert space.
\vskip .3cm \noindent
2000 \textit{Mathematics Subject Classification.} Primary 46N50, 47N50;
Secondary 46E15, 81S99

\section{Definitions and notation} 

   We begin with some definitions and notation.
We start with an introduction to $\mu$-deformed Segal-Bargmann analysis
(which is itself a realization of $\mu$-deformed quantum mechanics, though
we will not go into that here).
For background on these subjects, see \cite{MA} and \cite{RO}.
For recent related work, see \cite{AS}, \cite{PS1}, \cite{PS2},
\cite {PS3},
\cite{SISOL} and \cite{SOL}.
The introductions of \cite{PS1} and \cite{PS2}
provide more motivation for studying this topic.

First, we take $\mu > -1/2$ to be a fixed parameter throughout this article. 

\begin{definition}
\label{defmeasures}
Say $\lambda > 0$. 
We define measures in the complex plane  ${\bf C}$ by
\begin{gather*}
d\nu_{e,\mu,\lambda}(z) := \nu_{e,\mu,\lambda}(z) dx dy, \\
d\nu_{o,\mu,\lambda}(z) := \nu_{o,\mu,\lambda}(z) dx dy,  
\end{gather*}
whose densities are defined by
\begin{gather}
\label{defdene}
\nu_{e,\mu,\lambda}(z) := \lambda \frac{2^{ \frac{1}{2}-\mu }}
{\pi \Gamma(\mu+\frac{1}{2} )}
      K_{\mu-\frac{1}{2}} ( | \lambda^\frac{1}{2} z|^2 ) |
 \lambda^\frac{1}{2} z|^{2\mu +1}  \ , \\
\label{defdeno}
\nu_{o,\mu,\lambda}(z) := \lambda \frac{2^{ \frac{1}{2}-\mu }}{\pi \Gamma(\mu+\frac{1}{2} )}
       K_{\mu+\frac{1}{2}} ( | \lambda^\frac{1}{2} z|^2 ) |
       \lambda^\frac{1}{2} z|^{2\mu +1}   \ 
\end{gather}
for $0 \ne z \in {\bf C}$, 
where $\Gamma$  (the Euler gamma function) and $K_\alpha$  (the Macdonald function 
of order $\alpha$) are defined in \cite{LEB}. 
Moreover, $dxdy$ is Lebesgue measure in ${\bf C}$. 
\end{definition}

    The function $K_\alpha$ is also known as the modified Bessel function of the
third kind or Basset's function.
(See \cite{ER}, p.~5.)
But it is also simply known as a modified Bessel function. 
(See \cite{GRAD}, p.~961, and \cite{AB}, p. 374.) 
One way to identify the Macdonald function is to note the
following useful property: 
\begin{equation}
\label{kalpha}
            K_\alpha (x) = \int_0^\infty du \, e^{-x \, {\rm cosh} u} \, 
{\rm cosh} (\alpha u),
\end{equation}
for $x > 0$ and any $\alpha \in {\bf R}$.
(See \cite{LEB}, page 119.)
An explanation of how the Macdonald functions come into
this theory in a natural way is given in \cite{SOGTMP}.

    From the formulas (\ref{defdene}) and (\ref{defdeno}), 
one can see why the case $\mu = -1/2$ has not been included.
One should refer to the discussion of the Bose-like oscillator in \cite{RO}
(especially, note
Theorem~5.7) for motivation for the condition $\mu > -1/2$.

    Let ${\cal H} ({\bf C}) $ be the space of all holomorphic functions
$f : {\bf C} \rightarrow {\bf C}$.
We note that $f_e := (f + Jf)/2$ (respectively, $f_o := (f - Jf)/2$) defines the
even (respectively, odd) part of 
$f$, where $Jf(z):= f(-z)$ is the parity operator. So, $ f = f_e + f_o$. 

  We use throughout the article the standard notations for $L^p$ spaces and
their norms without further comment.
All $L^p$ spaces in this article are complex.
However, the ambiguous notation $ || \cdot ||_{p \rightarrow q} $ is used to denote the operator norm from
some $L^p$ space to some $L^q$ space without specifying
the measure spaces involved.
The context will indicate which measures spaces are meant.

\begin{definition}
\label{def2}
The \emph{$\lambda$-dilated, $\mu$-deformed
Segal-Bargmann space} defined for $0 < p < \infty$ and $\lambda > 0$  is 
\begin{equation*}
{\cal B}^p_{\mu,\lambda} := 
{\cal H} ({\bf C})  \cap \left\{ f : {\bf C} \rightarrow {\bf C} \ | \  
f_e \in L^p( {\bf C}, \nu_{e,\mu,\lambda}) {\rm ~and~}  f_o \in L^p( {\bf C}, \nu_{o,\mu,\lambda}) \right\},
\end{equation*}
where $ f = f_e + f_o$ is the decomposition of a function
into its even and odd parts.
Next we define
$$
|| f ||_{ {\cal B}^p_{\mu,\lambda} } :=
\left(  || f_e ||^p_{ L^p({\bf C},\nu_{e,\mu,\lambda} )} +
|| f_o ||^p_{ L^p( {\bf C}, \nu_{o,\mu,\lambda} )}
\right)^{1/p}
$$
for all $ f \in {\cal B}^p_{\mu,\lambda}$. 
We also define the {\em even subspace} of ${\cal B}^p_{\mu,\lambda}$ by
$$
    {\cal B}^p_{e,\mu,\lambda} := {\cal B}^p_{\mu,\lambda} \cap \{f : f = f_e\} 
$$
and the {\em odd subspace} of ${\cal B}^p_{\mu,\lambda}$ by 
$$
{\cal B}^p_{o,\mu,\lambda} := {\cal B}^p_{\mu,\lambda} \cap \{f : f = f_o\}. 
$$ 
\end{definition} 

        In these definitions we do not write the subscript $\lambda$ in the case
when $\lambda=1$.

  As far as we know, the two Definitions \ref{defmeasures} and \ref{def2}
are due to us but first appeared in print in joint work of the second
author with Pita in \cite{PS1}.
Behind these definitions there is a lot of history which we will relate
to the best of our knowledge.
As is customary, we do offer our sincerest apologies to those researchers whose
work we have not mentioned merely due to our own ignorance.
These definitions are due to the present authors in \cite{AS} in 2006 in the case
when $0 < p < \infty$ and $\lambda = 1$ and  $\mu > -1/2$
and to Marron \cite{MA} in 1994 in the case when $p=2$ and $\lambda > 0$
and $\mu > -1/2$.
However, Marron's work closely follows Rosenblum's in \cite{RO2} 
(also in 1994) where
the case $p=2$ and $\lambda=1$ and $\mu > -1/2$ is presented.
The works of Rosenblum and Marron were most influential for
our work on this topic.
However, in Sharma et al.~(\cite{SMMS}) formula (2.58) 
gives the inner product in equation (\ref{defip}) 
below up to a multiplicative constant.
So these authors already had in 1981 the case $p=2$ and $\lambda=1$
and $ \mu > -1/2$.
This is the earliest reference that we are aware of.
But slightly later in 1984 Cholewinski in \cite{CHO} has the case
$p=2$, $\lambda = 1$ and $\mu \ge 0$, but only for the
even subspace.
Next Sifi and Soltani in
\cite{SISOL} in 2002 have the case $p=2$, $\lambda = 1$ and $\mu \ge 0$.
Finally, we note that Ben~Sa\"id and {\O}rsted
in \cite{SBSBO} in 2006 present in detail
the case $p=2$ and $\lambda = 1$ and $\mu \ge 0 $
in Example 4.17, though they are aware of the case when $\mu$
is negative.

The next known result is elementary.
We include it here since it seems not to have been proved
in the literature before.

\begin{prop}
\label{prop11}
For $p \ge 1$ and $\lambda > 0$ 
we have that $ || \cdot ||_{ {\cal B}^p_{\mu,\lambda} }$ is a norm and that ${\cal B}^p_{\mu,\lambda}$  
is a Banach space which is the (internal) direct sum of the Banach subspaces 
${\cal B}^p_{e,\mu,\lambda}$  and ${\cal B}^p_{o,\mu,\lambda}$.
\end{prop}

\vskip .4cm \noindent
{\bf Proof:}
The proofs that $ || \cdot ||_{ {\cal B}^p_{\mu,\lambda} }$ is a norm 
and that we have a direct sum are straightforward and left to the reader.
It remains for us here to show that this space is complete.
This argument is well known
(for example, see  \cite{HA2}), and we give a sketch of it.

We first note that by definition $f \in {\cal B}^p_{\mu,\lambda}$
is equivalent to these conditions:
\begin{enumerate}

\item $f_e$ and $f_o$ are holomorphic in ${\bf C}$.

\item $f_e \in L^p ( {\bf C}, \nu_{e,\mu,\lambda} )$ and
$f_o \in L^p ( {\bf C}, \nu_{o,\mu,\lambda} )$.

\end{enumerate}
Since $f_e$ is holomorphic, we have by
the theory of a complex variable that
$$
         f_e(z) = \frac{1}{\pi r^2} \int_{B_r(z)} d\mu_L(w) \, f_e(w)
$$
for any $z \in {\bf C}$ and any $r>0$, where
$\mu_L$ is Lebesgue measure and
$B_r(z)$ is the ball of radius $r$ and center $z$.
So, using the fact that $ \nu_{e,\mu,\lambda}$ has no zeros, 
$$
f_e(z) = \frac{1}{\pi r^2} \int_{\bf C}
d\mu_L(w) \, \nu_{e,\mu,\lambda} (w) \left( 
\chi_{B_r(z)}(w) \frac{1}{\nu_{e,\mu,\lambda} (w)} \right) f_e(w)
$$
where $\chi_S$ denotes the characteristic function of a set $S$.
Applying H\"older's inequality, we get for all $z \in {\bf C}$ that
$$
|f_e(z)| \le C_e(z) || f_e ||_{ L^p(\nu_{e,\mu,\lambda})}
$$
where
$$
C_e(z)=\frac{1}{\pi r^2} \,\Big|  \Big| 
\frac{\chi_{B_r(z)}}{\nu_{e,\mu,\lambda}}\Big| \Big|
_{ L^{p^\prime}(\nu_{e,\mu,\lambda}) }
$$
is a finite real number that depends continuously on $z$.
Here $p^\prime$ is the usual dual Lebesgue index.
Similarly, we get
$$
|f_o(z)| \le C_o(z) || f_o ||_{ L^p(\nu_{o,\mu,\lambda})}
$$
where $ C_o(z)$ depends continuously on $z$.
To show that ${\cal B}^p_{\mu,\lambda}$ is complete, we take a
Cauchy sequence $f_n$ in that space and will show that it converges
to an element of the space.
But $f_n$ Cauchy in ${\cal B}^p_{\mu,\lambda}$ implies that
the sequence of even parts $(f_n)_e$ is Cauchy in $ L^p (\nu_{e,\mu,\lambda} )$
and that
the sequence of odd parts $(f_n)_o$ is Cauchy in $ L^p (\nu_{o,\mu,\lambda} )$.
Since $ p \ge 1$ these two Lebesgue spaces are complete and so
$ (f_n)_e \to g$ and $ (f_n)_o \to h$ as $n \to \infty$, where
$g \in  L^p (\nu_{e,\mu,\lambda} )$ and $h \in  L^p (\nu_{o,\mu,\lambda} )$.
Clearly, $g$ is even and $h$ is odd.
Now by a standard argument
the above two inequalities imply that $ (f_n)_e \to g$ and $ (f_n)_o \to h$
uniformly on compact subsets of ${\bf C}$, and so $g$ and $h$ are holomorphic.
This implies that $ g+h \in {\cal B}^p_{\mu,\lambda}$ and that
$ f_n \to g+h$ in the norm of ${\cal B}^p_{\mu,\lambda}$.
QED.
\vskip .4cm

Moreover, for $p=2$ we have that ${\cal B}^2_{\mu,\lambda}$ 
is a Hilbert space (see \cite{MA}) with inner product defined by 
\begin{equation} 
\label{defip}
\langle f, g \rangle_{ {\cal B}^2_{\mu,\lambda} } 
:= \langle f_e, g_e \rangle_{L^2(\nu_{e,\mu,\lambda})} + \langle f_o, g_o \rangle_{L^2(\nu_{o,\mu,\lambda})}.
\end{equation} 
Of course, $f = f_e + f_o$ and $g = g_e + g_o$ are the representations of $f$ and $g$ 
as the sums of their even and odd parts. 
(We will often use such representations without explicit comment, letting the notation 
carry the burden of explanation.) 
In this case, $ {\cal B}^2_{\mu,\lambda} $ is the Hilbert space (internal) direct sum
of the subspaces $ {\cal B}^2_{e,\mu,\lambda} $ and $ {\cal B}^2_{o,\mu,\lambda} $.
As we shall see in Section~3, each of the spaces $ {\cal B}^2_{\mu,\lambda} $, $ {\cal B}^2_{e,\mu,\lambda} $ 
and $ {\cal B}^2_{o,\mu,\lambda} $ is a reproducing kernel Hilbert space. 
When $\mu = 0$ and $\lambda=1$ 
this reduces to the usual Segal-Bargmann space, denoted here by $ {\cal B}^2$. 
(See \cite{BA} and \cite{SEG}.)
Further motivation for the nomenclature in Definition~\ref{def2} is given in \cite{SOGTMP}. 

        Note that $\nu_{e,\mu,\lambda}(z) = \lambda \nu_{e,\mu}(\lambda^{1/2} z)$ and 
$\nu_{o,\mu,\lambda}(z) = \lambda \nu_{o,\mu}(\lambda^{1/2} z)$, so that $\lambda$ is a
dilation parameter. 
Or, in other words, the dilation operator $T_\lambda$ defined by 
\begin{equation} 
\label{deftlambda}
T_\lambda f (z) := f (\lambda^{1/2} z)
\end{equation}
for $f \in {\cal B}^2_{\mu}$ and $ z \in {\bf C}$ 
is a unitary transformation from ${\cal B}^2_{e,\mu}$ onto ${\cal B}^2_{e,\mu,\lambda}$ and from 
${\cal B}^2_{o,\mu}$ onto ${\cal B}^2_{o,\mu,\lambda}$. 
Therefore,  $T_\lambda$
is also a unitary map from ${\cal B}^2_{\mu}$ onto ${\cal B}^2_{\mu,\lambda}$. 

   One can relate the parameter $\lambda$ to Planck's constant $\hbar$ by considering
the case $\mu =0$. 
We first observe that for $z \in {\bf C}$, $z \ne 0$ and $\mu =0$ we have that
$$
   \nu_{e,0,\lambda}(z) =  \nu_{o,0,\lambda}(z) = 
   \lambda \frac{2^{1/2}}{\pi \Gamma(1/2)} K_{1/2} ( |\lambda^{1/2} z|^2 ) \cdot |\lambda^{1/2} z| =
   \frac{\lambda}{\pi} e^{- \lambda |z|^2},
$$
which is a normalized Gaussian, using
$K_{1/2}(x) = K_{-1/2}(x) = ( \pi/ (2x) )^{1/2} e^{-x}$.
(See \cite{LEB}, p.~110 and p.~112.)
This should be compared with the Gaussian 
\begin{equation}
\label{defgauss}
  \nu_{ {\rm Gauss}, \hbar} (z) := \frac{1}{\pi \hbar} e^{- |z|^2 / \hbar}, 
\end{equation} 
which is the density for the measure of the Segal-Bargmann space for any $\hbar >0$. 
(See \cite{HA2}, p.~9 and p.~21. Note that the identification $t=\hbar$ is made in \cite{HA2}.) 
So it turns out that $\lambda = 1/\hbar $. 
(For those who are confused by the fact that $\hbar$ and $ |z|^2 $ have the same dimensions, 
let us note that there is a \emph{normalized} harmonic oscillator Hamiltonian 
implicitly used here. 
So there is both a mass and a frequency which have been taken equal to 
the dimensionless constant $1$.) 

\begin{definition} 
Let $(\Omega, \nu )$ be a measure space with finite measure (meaning that  $0 < \nu(\Omega) < \infty$).
Define the {\em entropy} 
of any $f$ in $L^2 (\Omega, \nu)$ to be 
\begin{equation} 
\label{defentropy}
    S_{L^2 (\Omega, \nu)} (f) := 
    \int_{\Omega} d\nu (\omega) |f(\omega)|^2 \log |f(\omega)|^2 - ||f||_{L^2 (\Omega, \nu)}^2 \log ||f||_{L^2 (\Omega, \nu)}^2,
\end{equation}
where $|| \cdot ||_{L^2 (\Omega, \nu)}$ means the norm in the Hilbert space $L^2 (\Omega, \nu)$, $\log$ is the
natural logarithm, and $0 \log 0 := 0$ (to make the function $0 \le r \mapsto r \log r$ 
continuous from the right at $r=0$). 
\end{definition}

This definition is due to Shannon \cite{SH} in his theory of communication. 
The requirement that the measure be finite
is not necessary, but is imposed to avoid technical details 
which are not important for us, since all the measure spaces in this article have finite measure. 
(See \cite{HI} for an example where $\nu(\Omega) = \infty$.) 
For a finite measure space we have that $S_{L^2 (\Omega, \nu)} (f)$ is defined for \emph{all}
$f \in L^2 (\Omega, \nu)$ and moreover that 
$$ 
(- \log W ) ||f ||^2_{L^2 (\Omega, \nu)} \le S_{L^2 (\Omega, \nu)} (f),
$$ 
where 
$W = \nu(\Omega)$, by applying Jensen's inequality to the probability space $(\Omega, \nu/W)$  
and the convex function $ r \mapsto r \log r$ for $r \ge 0$.
It follows that $S_{L^2 (\Omega, \nu)} (f) > -\infty$, though
$S_{L^2 (\Omega, \nu)} (f) = + \infty$ could occur.

\begin{definition}
\label{def14}
If there is a distinguished quadratic form $Q(f)$ defined
for all $f \in X$, a closed subspace of ${L^2(\Omega,\nu)}$ where 
$(\Omega,\nu)$ is a measure space, we say that
an inequality holding for all $f \in X$ of the form 
$$
                S_{L^2(\Omega,\nu)} (f) \le C_1 Q(f) + C_2 || f ||_{L^2(\Omega,\nu)}^2,
$$
for constants $C_1 >0$ and $C_2 \ge 0$ is a \emph{(direct) log-Sobolev inequality} in $X$.
Similarly, an inequality holding for all $f \in X$ of the form 
$$
          Q(f)  \le D_1 S_{L^2(\Omega,\nu)} (f) + D_2 || f ||_{L^2(\Omega,\nu)}^2,
$$
for constants $D_1 >0$ and $D_2 \ge 0$ is a \emph{reverse log-Sobolev inequality} in $X$.
\end{definition} 

Usually, $Q(f)$ in this definition is a Dirichlet form, but
this is not so in the main results given in Section~4.
We understand $Q: X \rightarrow [0,\infty]$ as a sort of energy. 
If $Q(f)$ is only densely defined, we put $Q(f) = +\infty$ for $f$ not in the
original domain of $Q$. 
Also, the entropy in this definition can be equal to $+\infty$.
So, one way to think about a direct log-Sobolev inequality is that it tells us that finite energy 
implies finite entropy. 
It can also be thought of as a type of coercivity inequality. 
Similar comments apply to reverse log-Sobolev inequalities.
There is a extensive literature on log-Sobolev inequalities,
starting with the articles
\cite{FE} of Federbush and \cite{GR} of Gross. 
For more recent references, see \cite{GZ} and references therein. 
The first reverse log-Sobolev inequality appeared in \cite{RLSI}. 
Further studies of such inequalities can be found in  \cite{AS},
\cite{BG}, \cite{GGS},
\cite{GG} and \cite{SOME}.

    We use the standard convention in analysis that $C$ represents a 
positive, finite constant (i.e., a quantity
not depending on the variable of interest in the context) which may change value 
with each usage. 

    The organization of the article is as follows. 
In Section~2, we review some basic properties of the measures introduced above. 
In Section~3 we analyze each reproducing kernel
function of the various Hilbert spaces 
studied here as the kernel function of an integral transform. 
In Section~4, we present our main result, an energy-entropy inequality
which in special
cases is a direct log-Sobolev inequality and in other cases is a reverse
log-Sobolev inequality. 
All of this is in terms of a quadratic form called the
$\mu$-deformed energy and introduced by the authors in \cite{AS}.
Then in Section~5 we present relations between the $\mu$-deformed energy 
and the Dirichlet form energy. 
This allows us to prove all of our main inequalities in terms of the Dirichlet form energy as well as in terms of the $\mu$-deformed energy.

\section{Properties of the measures}

We note the following results (see \cite{LEB}, p.~136) 
for the asymptotic behavior of the Macdonald function $K_\alpha(x)$ for
$\alpha \in {\bf R}$ and $x >0$:
\begin{gather*}
K_\alpha(x) \cong \frac{ 2^{|\alpha|-1} \Gamma(|\alpha|) }{ x^{|\alpha|} } \quad {\rm as} \quad x \rightarrow 0^+ 
\quad {\rm if} \quad \alpha \ne 0. \\
K_0(x) \cong  \log \frac{2}{x}   \quad {\rm as} \quad x \rightarrow 0^+ . \\
K_\alpha(x) \cong \left( \frac{ \pi }{ 2x } \right)^{1/2} e^{-x} \quad {\rm as} \quad x \rightarrow +\infty 
\quad {\rm for~all} \quad \alpha \in {\bf R}.
\end{gather*} 
Here $f(x) \cong g(x)$ as $x \rightarrow a$ 
means $\lim_{x \rightarrow a} f(x) / g(x) =1$, where  
$a$ is a limit point of a common domain of definition of the 
positive functions $f$ and $g$. 
While the usual definition of $K_\alpha$ (see \cite{LEB}, pp.~108-109) gives an analytic function 
defined on $ {\bf C} \setminus ( - \infty, 0 ]$, we are only interested in its values for real $ x > 0$. 
Notice that the asymptotic behavior of $K_\alpha(x)$ as $x \rightarrow +\infty$ does not depend on $\alpha$ 
to first order. But the next order term does depend on $\alpha$. 

    Written in polar coordinates $d\nu_{o,\mu,\lambda}$ has density (with respect to  $dr d\theta$) 
$$
\lambda^{\mu+\frac{3}{2}} \frac{2^{ \frac{1}{2}-\mu }}{\pi \Gamma(\mu+\frac{1}{2} )}
                      K_{\mu+\frac{1}{2}} (  \lambda r^2 ) r^{2\mu +2} .
$$
So, the behavior of the density of $d\nu_{o,\mu,\lambda}$ near zero ($r \rightarrow 0^+$) is asymptotic to 
$$
                  C \frac{1}{(r^2)^{(\mu+1/2)}} r^{2\mu +2} = C r 
$$
for all $\mu > -1/2$. 
On the other hand, $d\nu_{e,\mu,\lambda}$ has density 
$$
\lambda^{\mu+\frac{3}{2}} \frac{2^{ \frac{1}{2}-\mu }}{\pi \Gamma(\mu+\frac{1}{2} )}
                      K_{\mu-\frac{1}{2}} (  \lambda r^2 ) r^{2\mu +2} 
$$
in polar coordinates (again with respect to $dr d\theta$), 
whose asymptotic behavior as $ |z| = r \rightarrow 0^+$ is given by the following three cases: 

\begin{itemize}

\item[] 
1: For $ -1/2 < \mu < 1/2$, we have 
$\nu_{e,\mu,\lambda} (z) \cong C  
 r^{2\mu + 2} / (r^2)^{ ( \frac{1}{2} -\mu ) }   = Cr^{4\mu + 1}$.

\item[] 
2: For $ \mu = 1/2$, we have 
$\nu_{e,\mu,\lambda} (z) \cong C ( | \log r^2 | ) r^3 = C r^3 | \log r |$. 
Note that this is {\em not} the limit when $\mu \uparrow 1/2$ of the previous case. 

\item[] 
3: For $ \mu > 1/2$, we have 
$\nu_{e,\mu,\lambda} (z) \cong C   r^{2\mu + 2}  / (r^2)^{ (\mu - \frac{1}{2} ) } = C r^3$. 
So for this range of values of $\mu$, the functional form of the asymptotic
dependence on $r$ (for $r$ near zero) is independent of $\mu$, namely $r^3$,
though the constant does depend on $\mu$.
Also, this functional form 
is the limit when $\mu \uparrow 1/2$ of the first case. 

\end{itemize} 

   Note that in all cases the singularity of the Macdonald function at zero in the formulas 
(\ref{defdene}) and (\ref{defdeno}) has been regularized into a locally integrable function of $r$ 
near $r=0$ by the factor $r^{2\mu+2}$, which comes from a factor of $r^{2\mu+1}$ given in 
the definition of the densities of the measures and another factor of $r$ that comes from the change 
of variables $dx dy = r dr d\theta$. 

Using (\ref{kalpha}) we see immediately that 
$ |\alpha| < |\beta|$ implies that $K_\alpha (x) < K_\beta (x)$ 
for all $x > 0$.
In particular, we have $K_{\mu - 1/2}(x) < K_{\mu + 1/2}(x)$ for all $ x > 0$ 
provided that $|\mu - 1/2| < |\mu + 1/2|$.
But this last condition is equivalent to $\mu > 0$. 
So, for all $\mu > 0$ and all $z \in {\bf C}$ with $z \ne 0$ we have that
\begin{equation}
\label{eleo}
        \nu_{e,\mu,\lambda}(z) < \nu_{o,\mu,\lambda}(z). 
\end{equation}
In the case $ \mu = 0 $, we have already seen that 
$\nu_{e,0,\lambda}(z) = \nu_{o,0,\lambda}(z) $. 
Finally, in the case $ -1/2 < \mu < 0$
we have $K_{\mu + 1/2}(x) < K_{\mu - 1/2}(x)$ for all $ x > 0 $ 
since $| \mu + 1/2 | < | \mu - 1/2 | $, and so it follows for $0 \ne z \in {\bf C}$ that 
\begin{equation}
\label{olee}
  \nu_{o,\mu,\lambda}(z) < \nu_{e,\mu,\lambda}(z). 
\end{equation}

   Since $\nu_{e,\mu,\lambda}(z)$ and $\nu_{o,\mu,\lambda}(z)$ are integrable
near zero, continuous in ${\bf C} \setminus \{ 0 \}$ 
and decay as 
$r = |z| \rightarrow +\infty$ as $C r^{2\mu+1} e^{-\lambda r^2}$ 
(density with respect to $dr d\theta$), it follows that 
the measures $d\nu_{e,\mu,\lambda}(z)$ and $d\nu_{o,\mu,\lambda}(z)$ are finite. 
   It turns out that $d\nu_{e,\mu,\lambda}(z)$ is a probability measure. 
To show this we will use the identity  
$$
            \frac{d}{dx} \left[ x^\alpha K_\alpha (x) \right] = - x^\alpha K_{\alpha-1} (x) 
$$ 
for $ x > 0$. 
(See \cite{LEB}, p.~110.) 
So we now evaluate that 
\begin{eqnarray*} 
\nu_{e,\mu,\lambda}( {\bf C} ) = 
   \int_{ {\bf C} } d\nu_{e,\mu,\lambda}(z) &=& 
2 \pi \int_0^\infty dr \, \frac{2^{\frac{1}{2}-\mu}}{\pi \Gamma(\mu+1/2)} r^{2\mu+2} \lambda^{\mu + \frac{3}{2}}
K_{\mu-1/2} (\lambda r^2)  \\
&=& \frac{2^{\frac{1}{2}-\mu}}{\Gamma(\mu+1/2)} 
\int_0^\infty ds \, s^{\mu + \frac{1}{2}} K_{\mu-1/2} (s)  \\
&=& \frac{2^{\frac{1}{2}-\mu}}{\Gamma(\mu+1/2)} 
\int_0^\infty ds  \frac{d}{ds} \left( - s^{\mu + \frac{1}{2}} K_{\mu+1/2} (s) \right) \\
&=& \frac{2^{\frac{1}{2}-\mu}}{\Gamma(\mu+1/2)} \left( - s^{\mu + \frac{1}{2}} K_{\mu+1/2} (s) \right) 
\biggm|_0^\infty \\
&=& \frac{2^{\frac{1}{2}-\mu}}{\Gamma(\mu+1/2)} 2^{\mu - \frac{1}{2}} \Gamma(\mu+1/2) =1, 
\end{eqnarray*} 
where we used the definition of the measure $d\nu_{e,\mu,\lambda}(z)$, a change of variables, 
the above quoted identity, the fundamental theorem of calculus and 
the asymptotic behavior of $K_{\mu+1/2}$ at zero and at infinity. 
(Another way of thinking about this fact is given in \cite{SOGTMP}.) 
It now follows from (\ref{eleo}) or (\ref{olee}) that $d\nu_{o,\mu,\lambda}(z)$ is \emph{not} a 
probability measure when $\mu \ne 0$. 

The results of this article hold for every value of the scaling parameter $\lambda > 0$. 
However, to keep the notation manageable, we usually will put $\lambda = 1$ hereafter. 
Of course, the case of general $\lambda$ is implied by the case $\lambda = 1$ 
by applying a dilation. 

\section{The reproducing kernel and its \\ associated integral transform}

    There is a reproducing kernel function $K$ for ${\cal B}^2_\mu$
(see \cite{MA} and \cite{SBSBO}), which satisfies the usual reproducing property, namely,
$\langle K( \cdot, w)^*, f \rangle_{{\cal B}^2_\mu} = f(w)$ for all $f \in {\cal B}^2_\mu$ 
and $w \in {\bf C}$.
In fact, $K(z,w) = \exp_\mu (z^*w)$ for all $z,w \in {\bf C}$,
where the $\mu$-deformed exponential function (see  \cite{RO}) is
defined by $\exp_\mu (z) := \sum_{k=0}^\infty z^k / \gamma_\mu(k)$ and the 
$\mu$-deformed
factorial is defined recursively for all integers $k \ge 0$ by 
\begin{gather}
              \gamma_\mu(0) := 1          \quad {\rm and} \quad 
              \gamma_\mu(k) := (k + 2 \mu \chi_o (k) ) \gamma_\mu(k-1) {\rm ~~if~~} k \ge 1.
\label{defgamma}
\end{gather}
Finally, $\chi_o(k) = 0$ for $k$ even and $\chi_o(k) = 1$ for $k$ odd, that is, $\chi_o$ is nothing other 
than the characteristic function of the odd integers. 
Other conventions in force here are that $z^*$ is the complex conjugate of $z \in {\bf C}$
and that all inner products are anti-linear in the first argument and linear in the second.

    Notice that for the case $\mu =0$ we have $\gamma_0(k) = k!$ 
and so $\exp_0 (z) = e^z$.
In general, the idea is that for $\mu = 0$ we recover familiar objects and relations, while 
for $\mu \ne 0$ we obtain a deformation of the standard theory. 
But it can happen that the deformed theory $\mu \ne 0$ has properties identical to those 
in the case $\mu =0 $. 
For example, we have that $\gamma_\mu(0) = 1 = 0!$ and $\exp_\mu(0) = 1 = e^0$. 
See \cite{SOGTMP} for more details about this point of view. 

   The results of the following lemma are immediate consequences of these definitions. 
The proofs can be found in \cite{MA}. 
\begin{lemma} 
\label{expmulemma}
The function $\exp_\mu(z)$ satisfies the following properties:

\begin{enumerate} 

\item For all $\mu > -1/2$ the $\mu$-deformed exponential $\exp_\mu(z)$ 
is a holomorphic function whose domain is the entire complex plane ${\bf C}$, that is, 
it is an entire function. 

\item
For any $\mu > -1/2$ and all $z \in {\bf C}$ we have  
$ \left| \exp_\mu(z) \right| \le \exp_\mu ( |z| ) $. 

\item If $\mu \ge 0$, then we have 
$\left| \exp_\mu (z) \right| \le e^{|z|}$ for every $z \in {\bf C}$. 

\item If $ -1/2 < \mu <0$,  there is a $C_\mu > 0$ so that 
$\left| \exp_\mu (z) \right| \le C_\mu ( 1 +|z|^{|\mu|} ) e^{|z|}$ for every $z \in {\bf C}$.  

\end{enumerate} 

\end{lemma}

\begin{definition} 
For a measurable function $f = f_e + f_o : {\bf C} \rightarrow {\bf C}$ we now define 
an integral kernel transform, denoted $Kf$, that is associated 
to the reproducing kernel function $K$ for ${\cal B}^2_\mu$ as follows: 
\begin{equation} 
\label{ikt} 
  Kf(w) := 
                  \int_{\bf C} d\nu_{e,\mu}(z) K_e(z,w) f_e(z) 
               + \int_{\bf C} d\nu_{o,\mu}(z) K_o(z,w) f_o(z),
\end{equation}
provided both integrals converge absolutely, 
this being a restriction on $f$ as well as on $w \in {\bf C}$. 
\end{definition} 

      Here, of course, $K_e (z,w)$ and $K_o (z,w)$ refer to the even and odd parts of $K(z,w)$ with respect to 
the first variable $z$, 
and each is a kernel function for an integral kernel transform that enters in the definition (\ref{ikt}) 
as well as in the subsequent definition (\ref{definekeko}).
Notice that the first integral in (\ref{ikt}), if it exists, gives an even function in $w$, while the second integral 
in (\ref{ikt}), if it exists, gives an odd function in $w$.
This property depends on the explicit form of $K(z,w)$. 

    If $f \in {\cal B}^2_\mu$, the right hand side of
definition (\ref{ikt}) reduces to
$\langle K( \cdot, w)^*, f \rangle_{{\cal B}^2_\mu} = f(w)$, that is,
$Kf = f$ in this case.
Of course, this remark is the motivation for this definition.

       The kernel function $K(z,w)$ appears in \cite{MA}, while all three  kernel functions 
$K(z,w)$, $K_e(z,w)$ and $K_o(z,w)$ appear in \cite{SBSBO}. 
(Note that explicit formulas for these reproducing kernels are
given in Example 4.17 in \cite{SBSBO}, and they appear to disagree
with our formulas given below.
But they are indeed equal to ours, as they must be.)

     Notice that $K_e (z,w) = \exp_{\mu,e} (z^*w)$ and that $K_o (z,w) = \exp_{\mu,o} (z^*w)$, where 
$\exp_{\mu,e}$ and $\exp_{\mu,o}$ are the even and odd parts, respectively, of $\exp_\mu$. 
Since $\exp_{\mu,e} (z^*w)$ (resp., $\exp_{\mu,o} (z^*w)$) as a function of $z$ is in 
$L^q(\nu_{e,\mu})$ (resp., $L^q(\nu_{o,\mu})$) for $ 1 \le q < \infty$ 
and any fixed $w \in {\bf C}$, (which is a consequence of Lemma~\ref{expmulemma}
and the previously cited asymptotic behavior 
of the Macdonald function near infinity),  it follows 
by H\"older's inequality that
$Kf(w)$ is well defined for every $w \in {\bf C}$ provided that 
$f_e \in L^{p_1} ( {\bf C}, \nu_{e,\mu} )$ 
and
$f_o \in L^{p_2} ( {\bf C}, \nu_{o,\mu} )$ 
for some 
$ 1 < p_{1} \le \infty$ and $ 1 < p_{2} \le \infty$. 
One can use Morera's Theorem to show that 
the resulting function $w \mapsto Kf(w)$ is holomorphic for all $w \in {\bf C}$. 

    When $p=2$ the spaces ${\cal B}^p_{e,\mu}$ and ${\cal B}^p_{o,\mu}$
introduced in Definition~\ref{def2} become
Hilbert spaces of holomorphic functions with reproducing kernel
functions given by  
$K_e (z,w) = \exp_{\mu,e} (z^*w)$ 
for ${\cal B}^2_{e,\mu}$
and $K_o (z,w) = \exp_{\mu,o} (z^*w)$ for ${\cal B}^2_{o,\mu}$, 
where $z,w \in {\bf C}$. 
These kernels then have associated integral transforms, given by
\begin{equation}
\label{defKe}
          K_e f(w):=  \int_{\bf C} d\nu_{e,\mu}(z) K_e(z,w) f(z) 
\end{equation}
and 
\begin{equation}
\label{defKo}
          K_o f(w):=  \int_{\bf C} d\nu_{o,\mu}(z) K_o(z,w) f(z)
\end{equation}
for measurable $ f : {\bf C} \to {\bf C}$,
provided the integrals converge absolutely. 
These two integral kernel transforms will be basic for our analysis. 

    Notice that we follow here the very common convention of using the same
symbol to denote
both a kernel function as well as its associated integral kernel transform.
We have already done this before in equation (\ref{ikt}).

   We also consider $K_e \oplus K_o$, which is defined 
for $w \in {\bf C} $ as 
\begin{equation} 
\label{definekeko} 
  (K_e \oplus K_o) ( f \oplus g ) (w) :=  K_e f(w) + K_o g(w)
\end{equation} 
where $f, g : {\bf C} \rightarrow {\bf C}$ are measurable functions, provided that 
both integrals in (\ref{defKe}) and  (\ref{defKo})  converge absolutely. 
Again, suitable integrability conditions on $f$ and $g$ guarantee that the integrals 
exist for all $w \in {\bf C}$ and, in that case, the resulting functions 
$K_e f$ and $K_o f$ are holomorphic in the entire complex plane. 
Moreover, note that 
$K_e \oplus K_o : L^2(\nu_{e,\mu}) \oplus L^2(\nu_{o,\mu}) \rightarrow {\cal B}^2_{e,\mu} 
\oplus {\cal B}^2_{o,\mu} = {\cal B}^2_{\mu}$ is the orthogonal projection in the Hilbert 
space of the domain onto the codomain, where the latter, ${\cal B}^2_{\mu}$, 
is included in the former, $L^2(\nu_{e,\mu}) \oplus L^2(\nu_{o,\mu})$, 
by the map $ f = f_e + f_o \mapsto f_e \oplus f_o$. 
Notice that $L^2(\nu_{e,\mu}) \oplus L^2(\nu_{o,\mu})$ is an external direct sum of Hilbert
spaces, while ${\cal B}^2_{e,\mu} \oplus {\cal B}^2_{o,\mu}$ is an internal direct sum of
Hilbert spaces. 

    For all $w \in {\bf C}$ we have the identity 
$$ 
    Kf(w) = (K_e \oplus K_o) (f_e \oplus f_o) (w) = K_e(f_e)(w) + K_o(f_o)(w). 
$$ 
So the study of $K_e \oplus K_o$ and of $K$ reduces to the study of $K_e$ and $K_o$. 

    Let us note in passing that, while the transforms defined in 
(\ref{ikt}) and (\ref{definekeko}) can be viewed 
as the sum of two integral transforms (each with respect to its own measure space), one can 
easily rewrite these as \emph{one} integral transform with respect to 
the measure space $({\bf C} \times {\bf Z}_2, \nu_\mu)$, where ${\bf Z}_2 = \{ -1, +1 \}$ is a
multiplicative group, $\nu_\mu \vert_{ {\bf C} \times \{ +1 \} } := \nu_{e,\mu}$ 
and $\nu_\mu \vert_{ {\bf C} \times \{ -1 \} } := \nu_{o,\mu}$. 
The group ${\bf Z}_2$ can be identified with the Coxeter group (see \cite{SBSBO} and \cite{MR}) 
of this formalism. 

  Now a natural problem is to identify all quadruples $p_1$, $q_1$, $p_2$, $q_2$
of Lebesgue indices such that 
\begin{equation} 
\label{quadproblem} 
K_e \oplus K_o : L^{p_1}(\nu_{e,\mu}) \oplus L^{p_2}(\nu_{o,\mu}) \rightarrow 
{\cal B}^{q_1}_{e,\mu} \oplus {\cal B}^{q_2}_{o,\mu}
\end{equation}
is bounded, that is, the
operator norm with respect to the indicated domain 
and codomain is finite. 
And given that this operator is bounded, another problem is to ascertain if it is compact. 
For example, if $p_1 = q_1 = 2$ and $p_2 = q_2 = 2$, then $K_e \oplus K_o$ is bounded 
(since it is an orthogonal projection), but is not compact (since it 
is an orthogonal projection with infinite dimensional range). 

For the purposes of the present
exposition it is better to start with the more general problem of
identifying those quadruples $p_1, q_1, p_2, q_2$ for which $K_e \oplus K_o$ is a
bounded (or compact) transformation of  
$L^{p_1} (\nu_{e,\mu}) \oplus L^{p_2} (\nu_{o,\mu})$ to  
${\cal B}^{q_1}_{e,\mu,a_1} \oplus {\cal B}^{q_2}_{o,\mu,a_2}$ for some reals
$a_1$ and $a_2$.
So we wish to study when
\begin{equation} 
\label{wquadproblem} 
K_e \oplus K_o : L^{p_1}(\nu_{e,\mu}) \oplus L^{p_2}(\nu_{o,\mu}) \rightarrow 
{\cal B}^{q_1}_{e,\mu, a_1} \oplus {\cal B}^{q_2}_{o,\mu, a_2}
\end{equation}
is bounded or compact.

    Here we are using a weighted modification of the previously defined spaces.
Specifically,
\begin{gather}
{\cal B}^q_{e,\mu,a} := {\cal H} ({\bf C})  \cap \left\{ f : {\bf C} \rightarrow {\bf C} \ | \  
                                                                                      f = f_e \in L^q( {\bf C}, \nu_{e,\mu,a})  
                           \right\}, \nonumber \\
{\cal B}^q_{o,\mu,a} := {\cal H} ({\bf C})  \cap \left\{ f : {\bf C} \rightarrow {\bf C} \ | \  
                                                                                    f = f_o \in L^q( {\bf C}, \nu_{o,\mu,a}) 
                           \right\}, \nonumber
\end{gather} 
where $a \in {\bf R}$ and 
\begin{equation}
\label{wmeasdef}
                        d \nu_{e,\mu,a}(z) := e^{-a|z|^2} d\nu_{e,\mu}(z)  {\rm ~and~}   
                        d \nu_{o,\mu,a}(z) := e^{-a|z|^2} d\nu_{o,\mu}(z) .
\end{equation}
Notice that we allow the possibility here that $a$ is negative.
When $a=0$ we recover the spaces of Definition \ref{def2}
for the case $\lambda = 1$.
Strictly speaking, the notation for the measures defined in (\ref{wmeasdef})
conflicts with the notation of Definition \ref{defmeasures}, but we use it to avoid
even more complicated notation.
The point is that in the notation of the measures $d \nu_{e,\mu,\lambda}(z)$
and $d \nu_{o,\mu,\lambda}(z)$ in
Definition \ref{defmeasures} the variable $\lambda>0$ could be interpreted
as the variable $a \in \mathbb{R}$ in the measures in (\ref{wmeasdef}).
However, the measures in Definition \ref{defmeasures} are dilations
of the measures $d \nu_{e,\mu,1}(z) \equiv d \nu_{e,\mu}(z) $
and $d \nu_{o,\mu,1}(z) \equiv d \nu_{o,\mu}(z)$ (as we noted earlier),
while the measures
in (\ref{wmeasdef}) are given by a simple weight function (depending on
the parameter $a$) times the measures $d \nu_{e,\mu}(z) $ and $d \nu_{o,\mu}(z)$,
which do not depend on $a$.
It follows that the measures in (\ref{wmeasdef}) are not those of
Definition \ref{defmeasures} when $\mu \ne 0$.
However, for $\mu = 0$ the measures in (\ref{wmeasdef}) are related to
those of Definition \ref{defmeasures} by
$e^{-a|z|^2} d\nu_{e,0}(z) = \lambda^{-1}d \nu_{e,0,\lambda}(z)$,
where $\lambda = 1+a$ provided that $a \ne -1$.
(In our applications we always have $a > -1$.
See for example Theorem \ref{thm31} below.)

   Of course, this problem naturally splits into two problems, since the first (resp., second)
summand on the left side of (\ref{wquadproblem}) maps to the first (resp., second)
summand on the right side of (\ref{wquadproblem}). 
An answer is given in the following theorem.
\begin{theorem}
\label{thm31}
Let $ 1 < p \le \infty $ and $ 1 \le q < \infty$. 
Then for any  $a > p^\prime q / 4 -1$, the integral kernel transform $K_e$ 
(respectively, $K_o$) is a 
compact (and, hence, bounded) operator from 
$L^p (\nu_{e,\mu})$ to  ${\cal B}^q_{e,\mu,a}$ 
(respectively, from $L^p (\nu_{o,\mu})$ to  ${\cal B}^q_{o,\mu,a}$). 

    Consequently, for $ 1 < p_j \le \infty$ and $ 1 \le q_j < \infty$ 
and $a_j > p_j^\prime q_j / 4 -1$
for $j=1,2$ we have that 
$K_e \oplus K_o$ is a 
compact (and hence, bounded) operator from 
$L^{p_1} (\nu_{e,\mu}) \oplus L^{p_2} (\nu_{o,\mu})$ to \ 
${\cal B}^{q_1}_{e,\mu,a_1} \oplus {\cal B}^{q_2}_{o,\mu,a_2}$. 
\end{theorem}
Here, $p^\prime$ is the usual index conjugate to $p$, namely, $p^\prime = p /
(p-1) $ for $ 1 < p < \infty$ and $\infty^\prime = 1$ and $1^\prime = \infty $.

\vskip .3cm \noindent
{\bf Proof:} 
The proof is given for the case of $K_e$, since the other case of $K_o$ has a quite similar proof. 
(However, occasional parenthetical comments are given about the latter case.)
The tool to prove this result is the \emph{Hille-Tamarkin norm}. 
(See \cite{HT} and \cite{JO}.) 
For the Lebesgue indices $ 1 < p \le \infty $ and $ 1 \le q < \infty$ and the kernel function
$K_e$ this norm is given by
\begin{equation}
\label{htnorm}
||| K_e |||_{p,q} : = \left(
                                     \int_{\bf C} d\nu_{e,\mu,a} (w) 
                                      \left( 
                                               \int_{\bf C} d\nu_{e,\mu} (z) | K_e(z,w) |^{p^\prime}
                                      \right)^{q/p^\prime} 
                            \right)^{1/q}. 
\end{equation}
(For $K_o$, one has to use the measures $d\nu_{o,\mu,a}$ and $d\nu_{o,\mu}$.)

   In the following we continue to use the same symbol $K_e$ to represent 
the kernel function as well as the operator defined by that kernel function. 
The main property of the Hille-Tamarkin norm that will be used here is given next. 
(See \cite{HT} and \cite{JO}.)

\begin{itemize}

\item[]  If $||| K_e |||_{p,q}$ as given in (\ref{htnorm}) is finite, 
then the corresponding integral kernel transform $K_e$ is a compact operator and, hence, bounded 
from $L^p (\nu_{e,\mu})$ to  ${\cal B}^q_{e,\mu,a}$. 
Moreover, the operator norm from $L^p (\nu_{e,\mu})$ to  ${\cal B}^q_{e,\mu,a}$ is bounded 
above by the Hille-Tamarkin norm, namely,
$|| K_e ||_{p \rightarrow q} \le ||| K_e |||_{p,q}$. 
\end{itemize}

    We remark that the notation $ ||| K_e |||_{p,q} $ has the same
ambiguity as does $ || K_e ||_{p \rightarrow q} $, namely
that the relevant measures are omitted from the notation.
But again context will clarify this.

     So, the first step is to estimate the inner integral 
$\int_{\bf C} d\nu_{e,\mu} (z) | K_e(z,w) |^{p^\prime}$ 
in equation (\ref{htnorm}) 
in order to determine its dependence on $w$. 
To do this, note that we have the following estimate, which follows 
from the definition of the even part of a function and from
Lemma~\ref{expmulemma}: 
\begin{eqnarray} 
         | K_e(z,w)|  &=& \frac{1}{2} \left|  \exp_\mu(z^*w) + \exp_\mu(-z^*w)  \right|  \nonumber \\
                            &\le& \exp_\mu( |z| |w| ) 
                            \le  C_\mu ( 1 + |z|^{|\mu|}  |w|^{|\mu|} ) e^{|z| |w|}. 
\label{Kestimate} 
\end{eqnarray} 
We can take $C_\mu$ as in part~4 of Lemma~\ref{expmulemma}
for $-1/2 < \mu < 0$ and
$C_\mu =1$ for $\mu \ge 0$.
(The same estimate holds for $K_o$.)
In the following estimates, the reader should not confuse the kernel function $K_e$ with
the Macdonald function $K_{\mu-1/2}$.
Also, in agreement with our convention mentioned earlier,
the symbol $C$ in the following
is a positive finite constant (that is, independent of $w$, but not
necessarily of $\mu$ or $p$) which can change with each occurrence. 

      Using the estimate (\ref{Kestimate}), the definition of the measure $d\nu_{e,\mu}$, 
the asymptotics of the Macdonald function near $+\infty$ and a completion of the square, we have 
\begin{gather} 
     \int_{|z| \ge M} d\nu_{e,\mu} (z) | K_e(w,z) |^{p^\prime} 
    \le \, C \int_{|z| \ge M} d\nu_{e,\mu} (z) ( 1 + |z|^{|\mu|}  |w|^{|\mu|} )^{p^\prime} e^{p^\prime |z| |w|} \nonumber \\
=  \, C \int_M^\infty dr K_{\mu-1/2} (r^2) r^{2\mu + 2} ( 1 + r^{|\mu|}  |w|^{|\mu|} )^{p^\prime} e^{p^\prime r |w|} 
\nonumber \\
 \le C \int_M^\infty dr e^{-r^2} r^{2\mu + 1} ( 1 + r^{|\mu|}  |w|^{|\mu|} )^{p^\prime} e^{p^\prime r |w|}\nonumber \\
 = C e^{p^{\prime^2}  |w|^2 /4 } 
\int_M^\infty dr e^{-(r- p^\prime |w| / 2)^2} r^{2\mu + 1} ( 1 + r^{|\mu|}  |w|^{|\mu|} )^{p^\prime}.
\label{twointegrals} 
\end{gather} 
For our present purposes the particular value of $0 < M < \infty$ is not relevant. 
To estimate the integral in (\ref{twointegrals}), we first note that  
for any $\alpha \ge 0$ we have the estimate 
$$
              e^{-(x-\alpha)^2} \le e^{ \frac{1}{4} } e^\alpha e^{-x} 
$$
for all $ x \ge 0$, which can be shown by calculus. 
Also for any $r>0$ we have the elementary inequality
\begin{equation} 
\label{alphar} 
              ( 1 + \alpha )^r \le C ( 1 + \alpha^r)
\end{equation}
for all $\alpha \ge 0$, where $C$ depends only on $r$, and not on $\alpha$. 
Applying these two inequalities to the integral in (\ref{twointegrals}), we have that
\begin{gather*}
\int_M^\infty dr e^{-(r- p^\prime |w| / 2)^2} r^{2\mu + 1} ( 1 + r^{|\mu|}  |w|^{|\mu|})^{p^\prime}  \\
\le C  e^{p^\prime |w|/2} \int_M^\infty dr e^{-r} r^{2\mu + 1} ( 1 + r^{{p^\prime}|\mu|}  |w|^{{p^\prime}|\mu|} ) 
\le C  e^{p^\prime |w|/2} (1 + |w|^{{p^\prime}|\mu|} ). 
\end{gather*} 
Substituting this into (\ref{twointegrals}) we have that 
\begin{equation*}
\int_{|z| \ge M} d\nu_{e,\mu} (z) | K_e(w,z) |^{p^\prime} 
\le C e^{p^{\prime^2}  |w|^2 /4 }  e^{p^\prime |w|/2} (1 + |w|^{{p^\prime}|\mu|} ). 
\end{equation*}
Now we consider the case $|z| \le M$, for which we see that 
\begin{gather*}
\int_{|z| \le M} d\nu_{e,\mu} (z) | K_e(w,z) |^{p^\prime} 
\le C \int_{|z| \le M} d\nu_{e,\mu} (z) 
( 1 + |z|^{|\mu|}  |w|^{|\mu|} )^{p^\prime} e^{p^\prime |z| |w|} \\
\le  C \int_{|z| \le M} d\nu_{e,\mu} (z) 
( 1 + M^{|\mu|}  |w|^{|\mu|} )^{p^\prime}  e^{p^\prime M |w|} 
\le  C ( 1 + M^{|\mu|}  |w|^{|\mu|} )^{p^\prime}  e^{p^\prime M |w|} \\
\le  C ( 1 + |w|^{{p^\prime} |\mu|} )  e^{p^\prime M |w|},  
\end{gather*}
where we first used the estimate (\ref{Kestimate}), 
second applied $|z| \le M$ to the integrand, third estimated the integral by a constant, 
and finally used (\ref{alphar}) and then made an elementary estimate. 

    Putting all this together we have that 
\begin{eqnarray}
&&\int_{\bf C} d\nu_{e,\mu} (z) | K_e(w,z) |^{p^\prime} \nonumber \\ 
\label{c1c2} 
&\le& 
C \left( ( 1 + |w|^{{p^\prime} |\mu|} )  e^{p^\prime M |w|} + 
 e^{p^{\prime^2}  |w|^2 /4 }  e^{p^\prime |w|/2} (1 + |w|^{{p^\prime}|\mu|} ) \right). 
\end{eqnarray}
But now each of the terms of the right hand side (\ref{c1c2}) can obviously be bounded by 
$C \exp ( \beta p^{\prime^2}  |w|^2 )$ for any $ \beta >1/4 $ 
and all $w \in {\bf C}$, where now the
constant $C$ can depend on $\beta$ (as well as on $p$, $\mu$ and $M$), but not on $w$. 
So, the final estimate on the inner integral in (\ref{htnorm}) is 
$$
\int_{\bf C} d\nu_{e,\mu} (z) | K_e(w,z) |^{p^\prime} \le C e^{\beta p^{\prime^2}  |w|^2 }
$$
for {\em any} $\, \beta > 1/4$ and all $w \in {\bf C}$. 
Continuing with the computation of the Hille-Tamarkin norm of $K_e$
in equation (\ref{htnorm}), we have to take
the last expression to the power $q/p^\prime$ and then integrate with respect to 
the measure  $d\nu_{e,\mu,a} (w)$. 
(Using $d\nu_{o,\mu,a} (w)$ for $K_o$ gives the same results.) 
But this gives us the estimate 
\begin{gather}
||| K_e |||^q_{p,q} = 
   \int_{\bf C} d\nu_{e,\mu,a} (w) \left( \int_{\bf C} d\nu_{e,\mu} (z) | K_e(w,z) |^{p^\prime} \right)^{q/p^\prime} 
\nonumber \\
\le 
C \int_{\bf C} d\nu_{e,\mu,a} (w) e^{ \beta p^{\prime}q  |w|^2 } = 
C \int_0^\infty dr K_{\mu-1/2}(r^2) r^{2\mu +2} e^{-ar^2} e^{ \beta p^{\prime}q  r^2  }.
\nonumber
\end{gather}
Now this last integral converges if and only if it converges near infinity. 
But there it has the upper bound 
$$
C \int_{M^\prime}^\infty dr e^{-r^2}   r^{2\mu +1}   e^{-ar^2} e^{ \beta p^{\prime}q  r^2 },
$$
for some $M^\prime > 0$, which converges if and only if $ -1 -a + \beta p^{\prime}q < 0 $. 
This condition in turn is equivalent to $ a > \beta p^{\prime}q -1$. 
However, we have by hypothesis that $ a > p^{\prime}q / 4 -1$, which implies that 
we can pick {\em some} $ \beta > 1/4$ such that
$$
 a > \beta p^{\prime}q -1 > p^{\prime}q / 4 -1. 
$$ 
Using this value of $\beta$ in the above argument shows that 
$||| K_e |||_{p,q} < \infty$. The remaining assertions of the theorem now follow directly. 
QED.  

{\bf Remark:} 
The argument in this proof can be refined in the case $\mu > 0$ with the aim 
of getting an improved estimate for the Hille-Tamarkin norm and, hence, for the 
operator norm. 
Clearly, one can use part~3 of Lemma~\ref{expmulemma} (instead of part~4) in this case.
But we can use an even better estimate, that follows directly from (2.3.5) 
in \cite{RO}. 
This says that for all $ z \in {\bf C}$ and $\mu > 0$ we have that 
$ | \exp_\mu (z) | \le \exp_\mu ( Re(z) )$. 
However, we are not now trying to find optimal constants, nor do we believe it to be likely 
that the Hille-Tamarkin norm will produce them. 

\begin{corollary}
Let $ 1 < p \le \infty$ and $ 1 \le q < \infty$ be given with $p^\prime q < 4$. 
Then the integral kernel transform $K_e$ (respectively, $K_o$) is a compact 
(and hence bounded) operator from $L^p ( \nu_{e,\mu} )$ to $ {\cal B}^q_{e,\mu}$
(respectively,  from $L^p ( \nu_{o,\mu} )$ to $ {\cal B}^q_{o,\mu}$). 
\end{corollary}

{\bf Proof:} This is the special case $a = 0$ of the theorem. 
One only has to note that $ {\cal B}^q_{e,\mu,a} = {\cal B}^q_{e,\mu}$
and that $ {\cal B}^q_{o,\mu,a} = {\cal B}^q_{o,\mu}$ when $a=0$. 
QED. 

\vskip .4cm 
Theorem \ref{thm31} and its corollary generalize results proved in \cite{RKSBS}
for the case $\mu = 0$.
Notice that the relations $a > p^\prime q /4 -1$ of the theorem
and $ p^\prime q < 4$ 
of the corollary do not depend on the parameter $\mu$,
and so are identical with the
relations already found in \cite{RKSBS}.
However, the Hille-Tamarkin and operator norms most likely do depend on $\mu$, 
though only an analysis which calculates good lower bounds for these norms (or the 
norms themselves) can settle this question. 
Here we have presented only upper bounds. 
Also, notice that for the case $\mu =0$ it  is proved in \cite{RKSBS} that the integral 
kernel transform $K$ is unbounded if $ p^\prime q > 4$. 
It is reasonable to conjecture that this also holds for the case $\mu \ne 0$. 

\section{The main results}

To obtain the main results of this article 
we will use an interpolation theorem due to Stein. (See \cite{ST} or Theorem~3.6 in \cite{BE}.) 
This theorem is a generalization of the well known interpolation theorem of Riesz-Thorin. (See \cite{SW}.) 
The reason interpolation theory is used here is to obtain operator norm estimates 
that vary smoothly as the pair of Lebesgue indices varies.
This will allow us to take a derivative with respect
to the interpolation parameter $t$ as the reader will shortly see.
This derivative is central to the argument that we use.

Since the Stein theorem is not so widely known, we now quote it.
But first, let us  
recall that a {\it simple function} 
is a measurable function $f$ having a finite
range $R \subset {\bf C}$ such that $f^{-1}(z)$ is a set of finite measure 
for every $z \in R$, $ z \ne 0$.
\begin{theorem} {\rm (Stein \cite{ST})} 
Let $(\Omega_j,\nu_j)$ for $j=1,2$ be $\sigma$-finite measure spaces. 
Let $T$ be a linear transformation which takes 
simple complex-valued functions on $\Omega_1$ 
to measurable complex-valued functions on $\Omega_2$.  
Let $p_0$, $p_1$, $q_0$, $q_1$ be in $[ 1,\infty \, ]$. 
Then, for $ 0 \le t \le 1$, define $p_t$ and $q_t$ by 
$$ 
  p_t^{-1} = (1-t) p_0^{-1} + t p_1^{-1} ~~~and~~~
  q_t^{-1} = (1-t) q_0^{-1} + t q_1^{-1}. 
$$
Suppose that 
$u_0,u_1:\Omega_1 \rightarrow [0,\infty)$ and $k_0,k_1:\Omega_2 \rightarrow [0,\infty)$ 
are measurable functions such that 
for all simple $f: \Omega_1 \rightarrow {\bf C}$ we have 
\begin{eqnarray} 
|| \left( Tf \right) k_0 ||_{ L^{q_0} (\Omega_2, \nu_2) } &\le& 
                                     A_0 || fu_0||_{ L^{p_0}(\Omega_1, \nu_1) } \nonumber \\ 
and~~~~~
|| \left( Tf \right) k_1 ||_{ L^{q_1} (\Omega_2, \nu_2) } &\le&
                                     A_1 || fu_1||_{ L^{p_1}(\Omega_1, \nu_1) } 
\nonumber
\end{eqnarray}
for some finite constants $A_0 \ge 0 $ and $A_1 \ge 0 $. 
(Note that for some simple $f$ the right side of these inequalities can be equal to $+\infty$.)
For $ 0 \le t \le 1$, define functions 
$u_t:= u_0^{1-t} u_1^t : \Omega_1 \rightarrow [0, \infty)$ and
$k_t:=k_0^{1-t} k_1^t : \Omega_2 \rightarrow [0, \infty)$.
Then the transformation $T$ can be extended uniquely to a linear transformation defined on the space 
of all $f: \Omega_1 \rightarrow {\bf C}$ that satisfy 
$ ||f u_t ||_{ L^{p_t}(\Omega_1, \nu_1)} < \infty $ 
in such a way that for all such $f$ we have 
\begin{equation} 
|| \left( Tf \right) k_t ||_{ L^{q_t} (\Omega_2, \nu_2) } \le A_0^{1-t} A_1^t 
|| f u_t ||_{ L^{p_t} (\Omega_1, \nu_1)} . \nonumber
\end{equation}
\end{theorem} 

    Now we will apply Stein's Theorem in the context of Theorem~\ref{thm31}.
The next result, including its proof using Stein's Theorem, follows the
presentation in \cite{RKSBS} for the case $\mu=0$. 
Moreover, the next result and its proof are valid for $K_o$ provided 
that we change the subscript ``$e$'' to ``$o$'' throughout. 

\begin{theorem}
Let $1 < p \le \infty$, $1 \le q < \infty$ and $a > p^\prime q /4 -1$. 
Then we have $ || K_e ||_{p \rightarrow q} \le  ||| K_e |||_{p,q}  < \infty $
and also that
for all $ 0 \le t \le 1$, $K_e$ is a bounded linear map from
$L^{p_t} ( {\bf C}, \nu_{e,\mu})$ to $L^{q_t} ( {\bf C}, \nu^t_{e,\mu,a})$, 
where
$$
   d\nu^t_{e,\mu,a}(z) := \exp \left( - \frac{t q_t}{q} a |z|^2 \right)
   d\nu_{e,\mu}(z)
$$
for $p_t^{-1} = (1-t) 2^{-1} + t p^{-1}$ and $q_t^{-1} = (1-t) 2^{-1} + t q^{-1}$.
Moreover, the operator norm from
$L^{p_t} ( {\bf C}, \nu_{e,\mu})$ to $L^{q_t} ( {\bf C}, \nu^t_{e,\mu,a})$ satisfies 
$$
  || K_e ||_{p_t \rightarrow q_t} \le
\left(  || K_e ||_{p \rightarrow q} \right)^t
< \infty,
$$
or equivalently, 
\begin{equation}
\label{bound1}
 || \left( K_e f \right) k_t ||_{L^{q_t} \left(  \nu_{e,\mu} \right)} \le A_e^t || f ||_{L^{p_t} \left(  \nu_{e,\mu}   \right)}, 
\end{equation}
for all $ f \in L^{p_t} ( {\bf C}, \nu_{e,\mu})$,
where $A_e = A_e(p,q,a,\mu):= || K_e ||_{p \rightarrow q} < \infty$
is the operator norm from 
$L^{p} ( {\bf C}, \nu_{e,\mu})$ to $L^{q} ( {\bf C}, \nu_{e,\mu,a})$ and where
\begin{equation}
\label{ktdef} 
k_t(z) = \exp ( - at |z|^2 / q)
\end{equation}
for all $z \in {\bf C}$ and $ 0 \le t \le 1 $.
\end{theorem}
{\bf Proof:} In the context of Stein's theorem, we take 
$(\Omega_1, \nu_1) = ( {\bf C}, \nu_{e,\mu})$ and 
$(\Omega_2, \nu_2) = ( {\bf C}, \nu_{e,\mu,a})$. 
Also take $p_0 = q_0 = 2$, $p_1 = p$, $q_1 = q$ and $k_0(z) = u_0(z) = u_1(z) = 1$ 
for all $z \in {\bf C}$. 
Finally, put $k_1(z) = \exp ( - a |z|^2 / q )$. 
Note first off that 
$$
            || \left( K_e f \right) k_t ||_{ L^{q_t} \left (\nu_{e,\mu} \right)  } = 
            ||K_e f ||_{ L^{q_t} \left (\nu^t_{e,\mu,a} \right)   }. 
$$
Here $k_t(z) = k_0^{1-t} (z) k_1^t (z)$
comes from the statement of Stein's Theorem.
Using the definitions for $k_0(z)$ and $k_1(z)$ just given, we get that 
$k_t(z) = k_1^t (z) = \exp ( - at |z|^2 / q)$, which is just
equation (\ref{ktdef}).
Note that $k_t$ also depends on $a$ and $q$, although this is suppressed from the notation. 
For $t = 0$, we have 
$$
         || \left( K_e f \right) k_0 ||_{L^2 \left(  \nu_{e,\mu}   \right)} \le || f u_0 ||_{L^2 \left(  \nu_{e,\mu}   \right)}
$$
for all $f \in L^2 \left(  \nu_{e,\mu}   \right)$, since $K_e$ is an orthogonal projection when considered as an operator
with domain $L^2 \left(  \nu_{e,\mu}   \right)$.
For $t = 1$, we can apply Theorem~\ref{thm31} because of our hypotheses on 
$p$, $q$ and $a$ and so we have that
$$
 || \left( K_e f \right) k_1 ||_{L^q \left(  \nu_{e,\mu}   \right)} \le A_e || f u_1 ||_{L^p \left(  \nu_{e,\mu}   \right)}. 
$$
(Recall that $A_e = || K_e ||_{p \rightarrow q}$.)
So, Stein's Theorem allow us to conclude that 
$$
 || \left( K_e f \right) k_t ||_{L^{q_t} \left(  \nu_{e,\mu} \right)} \le 
    1^{1-t} A_e^t || f u_t ||_{L^{p_t} \left(  \nu_{e,\mu}   \right)} = A_e^t || f ||_{L^{p_t} \left(  \nu_{e,\mu}   \right)}, 
$$ 
or, equivalently, 
$$
 || K_e f ||_{L^{q_t} \left(  \nu^t_{e,\mu,a} \right)} \le A_e^t || f ||_{L^{p_t} \left(  \nu_{e,\mu}   \right)} 
$$ 
for all $f \in L^{p_t} ( \nu_{e,\mu} )$. 
Here we have used $u_t = u_0^{1-t} u_1^t \equiv 1$. QED.

  In the next theorem and its discussion we will see three
expressions arising quite naturally.
These have been basically
identified by us in \cite{AS} and are given next.
We give these definitions 
for the measures introduced in Definition~\ref{defmeasures}.
\begin{definition}
\label{defgene}
Let $\lambda >0$ be a given value throughout of the dilation parameter.
For every $ g \in {\cal B}^2_{e,\mu,\lambda}$ define its {\em $\mu$-deformed energy} by
\begin{equation}
\label{defeemu}
     E_{e,\mu,\lambda} (g) := \int_{ {\bf C} } d\nu_{e,\mu,\lambda} (z)  \, \lambda \, |z|^2 | g(z)|^2.
\end{equation}
Similarly, for every $ h \in {\cal B}^2_{o,\mu,\lambda}$ define its {\em $\mu$-deformed energy} by
\begin{equation*}
     E_{o,\mu,\lambda} (h) := \int_{ {\bf C} } d\nu_{o,\mu,\lambda} (z) \, \lambda \, |z|^2 | h(z)|^2.
\end{equation*}
Finally, for every $ f \in {\cal B}^2_{\mu,\lambda}$ define its {\em $\mu$-deformed energy} by
\begin{equation}
\label{defemu}
   E_{\mu,\lambda} (f) := E_{e,\mu,\lambda} (f_e) + E_{o,\mu,\lambda} (f_o),
\end{equation}
where $f = f_e + f_o$ is the
representation of $f$ as the sum of its even and odd parts. 
\end{definition}
See \cite{AS} for the case $\lambda =1$ of this definition.
With the normalization we have chosen, we have that
$E_{\mu,1} (f) = E_{\mu,\lambda} (T_\lambda f)$ 
for all $f \in {\cal B}^2_{\mu}$, where $T_\lambda$
is defined in equation (\ref{deftlambda}).
Having made this comment, we now revert to the situation where $\lambda = 1$
and $\lambda$ is suppressed from the notation. 

    We note that all of these $\mu$-deformed energies are non-negative
quantities, although they can be equal to $+\infty$. 
We have given in \cite{AS} explicit formulas for these 
$\mu$-deformed energies in terms of the
coefficients of the Taylor series (centered in the origin) of the function. 
Unfortunately, those formulas are rather unenlightening and do not show an immediate 
relation with the Dirichlet form energy, which we introduce in the next section. 
Note that in the case $\mu = 0$ these $\mu$-deformed energies are related to the Dirichlet 
energy in the Segal-Bargmann space ${\cal B}^2$ via an identity of Bargmann 
that is proved in \cite{BA} (equation (3.17)), namely, for all $f \in {\cal B}^2$ we have that 
$$
   \int_{ {\bf C} } d\nu_{\rm Gauss} (z) |z|^2 |f(z)|^2 = || f ||^2_{ {\cal B}^2 } + \left\langle f, Nf \right\rangle_{ {\cal B}^2 },
$$
where $d\nu_{\rm Gauss} \, ( \, =  d\nu_{e,0} = d\nu_{o,0} )$ is a Gaussian
measure (cp. equation (\ref{defgauss})) and $N$ is the
number operator which is associated with the Dirichlet form.
See \cite{BA} for more details.
In the next section
we will discuss a $\mu$-deformed number operator $N_\mu$
acting in ${\cal B}^2_\mu$
and its associated Dirichlet form as well as its relation 
with the $\mu$-deformed energies of Definition~\ref{defgene}.

    We now continue with the main results of this article. 

\begin{theorem}
\label{thm43}
Suppose that $1 < p \le \infty$, $1 \le q < \infty$ and $a > p^\prime q /4 -1$. 
Then the energy-entropy inequality
\begin{equation}
\label{eeineqe}
\left( p^{-1} - q^{-1} \right) S_{ L^2 \left( \nu_{e,\mu} \right) } \left( f \right) \le 
\left( \log A_e \right) || f ||^2_{ L^2 \left( \nu_{e,\mu} \right) } + \frac{a}{q}E_{e,\mu} (f) 
\end{equation} 
holds, where $A_e = A_e(p,q,a,\mu)$ is the operator norm of $K_e$ acting from $L^p (\nu_{e,\mu})$ 
to ${\cal B}^q_{e,\mu,a}$, provided that one of the following hypotheses is satisfied:
\begin{itemize} 

\item[] Hypothesis~1:  $ f \in {\cal B}^{2+\epsilon}_{e,\mu}$ for some $\epsilon > 0$.

\item[] Hypothesis~2: $ f \in {\cal B}^{2}_{e,\mu}$, $1 < p \le 2$, $1 \le q \le 2$ 
and $S_{L^2(\nu_{e,\mu})} (f) < \infty$.

\end{itemize} 
Moreover, for the coefficients of the principle terms 
in (\ref{eeineqe}), namely the energy term $E_{e,\mu} (f)$ and the entropy term 
$S_{ L^2 \left( \nu_{e,\mu} \right) } \left( f \right) $,  
we have the following cases: 

\begin{itemize}

\item[]
Case~1: $p^{-1} > q^{-1}$. This implies that $ p^\prime q /4 -1 >0$ and so $a > 0$. 
Thus the coefficients of both $S_{ L^2 \left( \nu_{e,\mu} \right) } \left( f \right) $ and 
$E_{e,\mu} (f)$ are positive and consequently (\ref{eeineqe}) is 
a \emph{direct log-Sobolev inequality} in ${\cal B}^{2}_{e,\mu}$ with 
respect to the $\mu$-deformed energy $E_{e,\mu}$.  

\item[]
Case~2: $p^{-1} \le q^{-1}$ and $p^\prime q /4 -1 \ge 0$. 
Again $a > 0$ follows so that 
the coefficient of $E_{e,\mu} (f)$ is positive, but now 
the coefficient of the entropy is non-positive. 
Since $d\nu_{e,\mu}(z)$ is a probability measure, $S_{ L^2 ( \nu_{e,\mu} ) } (f) \ge 0$ and 
so (\ref{eeineqe}) is trivially true. 

\item[]
Case~3: $p^\prime q /4 -1 < 0$. This implies that  $p^{-1} < q^{-1}$, namely, 
that the coefficient of the entropy is negative. 
Moreover, we choose $a$ such that $0 > a > p^\prime q /4 -1$, 
which means that the energy term also has a negative coefficient. 
(Of course, we can also choose $a \ge 0$ in this case.
But then (\ref{eeineqe}) becomes trivial.)
In this case by putting the energy term on the left and the entropy term on the right,
(\ref{eeineqe}) gives us a \emph{reverse log-Sobolev inequality}  in ${\cal B}^{2}_{e,\mu}$ 
with respect to the $\mu$-deformed energy $E_{e,\mu}$.

Since $K_e 1 = 1$ (where $1$ is the constant function, which is holomorphic and even), 
we have that $A_e \ge 1$ and so the coefficient
of the norm term in (\ref{eeineqe}) is non-negative. 
Here, we use that $a < 0$ implies $ ||1||_{ {\cal B}^q_{e,\mu,a} } \ge 1$. 

\end{itemize}

\end{theorem}

\vskip .4cm \noindent
{\bf Remark:} The corresponding inequality holds for odd functions. 
One merely has to change the subscript ``$e$'' to ``$o$'' throughout. 
We simply note the result here. 
So, with the same hypotheses as in Theorem~\ref{thm43}, we have that
\begin{equation}
\label{eeineqo}
\left( p^{-1} - q^{-1} \right) S_{ L^2 \left( \nu_{o,\mu} \right) } \left( f \right) \le 
\left( \log A_o \right) || f ||^2_{ L^2 \left( \nu_{o,\mu} \right) } + \frac{a}{q}E_{o,\mu} (f), 
\end{equation} 
where $A_o = A_o(p,q,a,\mu)$ is the operator norm of $K_o$ acting from $L^p (\nu_{o,\mu})$ 
to ${\cal B}^q_{o,\mu,a}$.
However, the comments about the three cases need some modification.
In Case~2 we remark that for $\mu >0$ we can have negative entropies and (\ref{eeineqo}) 
could be non-trivial for some choices of $f$.
Also, the second paragraph of Case~3 does not apply.

    The proof (in either the even or odd case) is essentially identical 
to that given in \cite{RLSI}, except for some notational changes 
some of which are due to the absence of a Bargmann identity for $E_{e,\mu} (f)$ 
and some to a difference in the normalization of the measures. 
Since the proof in \cite{RLSI} is rather long and technical, it will not be
repeated in detail here.
However, we now present a sketch of the main ideas of the proof. 

    We start with the formula (\ref{bound1}), which we repeat here: 
\begin{gather*}
\hspace{2.55cm}
 || \left( K_e f \right) k_t ||_{L^{q_t} \left(  \nu_{e,\mu} \right)} \le A_e^t || f ||_{L^{p_t} \left(  \nu_{e,\mu}   \right)}.  
\hspace{2.55cm}
(\ref{bound1}) 
\end{gather*}
This is valid with $A_e$ \emph{finite} because of the assumptions imposed on $p$, $q$ and $a$. 
We have proved this formula for $f \in L^{p_t}(\nu_{e,\mu})$ and 
hence for $f \in L^{p_t}(\nu_{e,\mu}) \cap {\cal B}^2_{e,\mu}$. 
But we will now use it for $f \in {\cal B}^2_{e,\mu}$. 
In the rest of this sketch, such technical details about domain issues
and their ensuing complications will be omitted. 
The idea is that  (\ref{bound1}) is an \emph{equality} when $t=0$, since $p_0 = q_0 = 2$, 
$k_0 \equiv 1$, $A_e^0 = 1$ and $K_e f = f$ because $f \in {\cal B}^2_{e,\mu}$. 
So, using a technique that dates back at least to 
Hirschman in \cite{HI} but that also is important in \cite{GR}, 
we take $f$, $p$ and $q$ fixed and regard each side 
of (\ref{bound1}) as a real-valued function of
the real variable $t \in [0,1]$.
The fact that equality obtains at $t=0$ implies that we can take the
derivative (from the right) at $t=0$ of both sides of (\ref{bound1})
and thereby get another valid inequality, namely
$$
      \frac{d}{dt} \biggm|_{t=0^+} \left(  ||  f  k_t ||_{L^{q_t} \left(  \nu_{e,\mu} \right)}  \right) 
\le \frac{d}{dt} \biggm|_{t=0^+} \left( A_e^t || f ||_{L^{p_t} \left(  \nu_{e,\mu}   \right)} \right),
$$
which simplifies to
\begin{equation}
\label{simp}
\frac{d}{dt} \biggm|_{t=0^+} \left( || f k_t ||_{L^{q_t} \left(  \nu_{e,\mu} \right)}  \right) 
\le ( \log A_e ) ||f||_{L^2(\nu_{e,\mu})} + 
\frac{d}{dt} \biggm|_{t=0^+} \left( || f ||_{L^{p_t} \left(  \nu_{e,\mu}   \right)} \right). 
\end{equation}
Note that derivation in general is \emph{not} an order preserving operator, but that in this 
particular instance, the operator $d/dt|_{t=0^+}$ is. 
Using elementary calculus, a differentiation under the integral sign (which we do not justify here) 
and the definition in equation (\ref{defentropy}) of entropy, we find that 
\begin{equation}
\label{dpt}
       \frac{d}{dt} \biggm|_{t=0^+} \left(  || f ||_{L^{p_t} \left(  \nu_{e,\mu}   \right)} \right) =
       \frac{ (2^{-1} - p^{-1}) S_{L^2(\nu_{e,\mu})}(f) }{||f||_{L^2(\nu_{e,\mu})}}, 
\end{equation} 
provided that $||f||_{L^2(\nu_{e,\mu})} \ne 0$. 
But (\ref{eeineqe}) is trivially true if $f \equiv 0$, so hereafter we exclude that case.
Similarly, we find that 
\begin{gather}
\frac{d}{dt} \biggm|_{t=0^+} \left( || f  k_t ||_{L^{q_t} \left(  \nu_{e,\mu} \right)}  \right) 
=  \frac{ (2^{-1} - q^{-1}) S_{L^2(\nu_{e,\mu})}(f) }{||f||_{L^2(\nu_{e,\mu})}}
\nonumber \\
\label{dqt}
+ \frac{1}{||f||_{L^2(\nu_{e,\mu})}} \int_{ {\bf C} } d\nu_{e,\mu}(z) \left( -\frac{a}{q}|z|^2 \right) |f(z)|^2, 
\end{gather}
using the following immediate consequence of equation (\ref{ktdef}):
$$
                \frac{dk_t}{dt} \biggm|_{t=0^+} = -\frac{a}{q} |z|^2.
$$
Substituting (\ref{dpt}) and  (\ref{dqt}) into (\ref{simp}) and using the definition 
in equation (\ref{defeemu})
of the $\mu$-deformed energy $E_{e,\mu}(f)$, we obtain
\begin{gather}
 \frac{ (2^{-1} - q^{-1}) S_{L^2(\nu_{e,\mu})}(f) }{||f||_{L^2(\nu_{e,\mu})}}
 - \frac{1}{||f||_{L^2(\nu_{e,\mu})}} \frac{a}{q} E_{e,\mu}(f) \nonumber \\
 \le 
 ( \log A_e ) ||f||_{L^2(\nu_{e,\mu})} + \frac{ (2^{-1} - p^{-1}) S_{L^2(\nu_{e,\mu})}(f) }{||f||_{L^2(\nu_{e,\mu})}}. 
\nonumber
\end{gather}
Then, multiplying by $||f||_{L^2(\nu_{e,\mu})}$, putting the two entropy 
terms on the left hand side and the energy term on the right hand side, we obtain (\ref{eeineqe}).
This concludes the sketch of the proof.
QED

As noted before, the missing details of the proof, which amount to some ten
pages, can be found in \cite{RLSI}.
It is in those details that Hypotheses~1 and~2 play a role in justifying
the differentiation under the integral sign, as mentioned earlier.

Now we state a corollary of a part of the proof that we have not
presented here.
Again, refer to \cite{RLSI} for details.
Notice that this is not a consequence of the conclusion of the previous theorem.

\begin{corollary} 
\label{maincor}
The following relations between entropies and $\mu$-deformed energies hold:
\begin{enumerate}
\item
For all $f_e \in {\cal B}^2_{e,\mu}$ we have that the Shannon entropy 
$S_{ L^2(\nu_{e,\mu})} (f_e)$ is finite if and only if the $\mu$-deformed energy  
$E_{e,\mu} (f_e)$ is finite. 
\item
For all $f_o \in {\cal B}^2_{o,\mu}$ we have that the Shannon entropy 
$S_{ L^2(\nu_{o,\mu})} (f_o)$ is finite if and only if the $\mu$-deformed energy  
$E_{o,\mu} (f_o)$ is finite.   
\item
For all $f \in {\cal B}^2_\mu$ we have that the $\mu$-deformed entropy
$S_{\mu}(f)$ (see Definition \ref{defmudeent} below)
is finite if and only if the $\mu$-deformed energy
$E_{\mu} (f)$ is finite.
\end{enumerate}
\end{corollary}

   By adding the inequalities (\ref{eeineqe}) and (\ref{eeineqo}) 
for the even and odd cases, we get the next result. 

\begin{corollary}
\label{corevenodd}
Let $1 < p_e \le \infty$, $1 \le q_e < \infty$, $a_e > p_e^\prime q_e /4 -1$, 
$1 < p_o \le \infty$, $1 \le q_o < \infty$ and $a_o > p_o^\prime q_o /4 -1$.
Then we have the energy-entropy inequality 
\begin{gather} 
\left( p_e^{-1} - q_e^{-1} \right) S_{ L^2 \left( \nu_{e,\mu} \right) } \left( f_e \right) 
+ 
\left( p_o^{-1} - q_o^{-1} \right) S_{ L^2 \left( \nu_{o,\mu} \right) } \left( f_o \right) 
\nonumber \\
\le \left( \log A_e \right) || f_e ||^2_{ L^2 \left( \nu_{e,\mu} \right) } 
+
\left( \log A_o \right) || f_o ||^2_{ L^2 \left( \nu_{o,\mu} \right) } 
+ 
\frac{a_e}{q_e}E_{e,\mu} (f_e) 
+
\frac{a_o}{q_o}E_{o,\mu} (f_o), 
\nonumber
\end{gather}
where $f = f_e + f_o$ is the representation of $f$ as the sum of its even and odd parts,
provided that $f_e$ (resp., $f_o$) satisfies 
one of the two Hypotheses of Theorem~\ref{thm43}
(resp., one of the two Hypotheses of Theorem~\ref{thm43}
with ``o'' instead of ``e''). 
\end{corollary}

\vskip .3cm \noindent
{\bf Remark:} 
We now will make a comparison of the present results with
our previous results in \cite{AS}. 
Note that Theorem~\ref{thm43}, Case~3, gives 
$$
             E_{e,\mu} (f_e) \le \frac{q}{a} ( p^{-1} - q^{-1} ) S_{ L^2 ( \nu_{e,\mu} ) } ( f_e ) 
         + \sigma_e (p,q,a,\mu) || f_e ||^2_{ L^2 ( \nu_{e,\mu} ) }
$$
for some constant $\sigma_e (p,q,a,\mu)$. 
It is shown by the second author in \cite{RLSI} that the coefficient of the entropy term can achieve any 
number $c > 1$. 
So we have:
\begin{theorem} 
(Reverse log-Sobolev inequalities in ${\cal B}^2_{e,\mu}$ and ${\cal B}^2_{o,\mu}$ for the $\mu$-deformed energy.) 
For every $f_e \in {\cal B}^2_{e,\mu}$ we have that 
\begin{equation}
\label{rlsie}
             E_{e,\mu} (f_e) \le c S_{ L^2 ( \nu_{e,\mu} ) } ( f_e ) 
         + \tau_e(c,\mu) || f_e ||^2_{ L^2 ( \nu_{e,\mu} ) }
\end{equation}
for any $c > 1$, where $\tau_e(c,\mu)$ is some finite constant. \\
For every $f_o \in {\cal B}^2_{o,\mu}$ we have that 
\begin{equation} 
\label{rlsio}
             E_{o,\mu} (f_o) \le c S_{ L^2 ( \nu_{o,\mu} ) } ( f_o ) 
         + \tau_o(c,\mu) || f_o ||^2_{ L^2 ( \nu_{o,\mu} ) }
\end{equation}
for any $c > 1$, where $\tau_o(c,\mu)$ is some finite constant.
\end{theorem}

The inequality (\ref{rlsie}) (resp., (\ref{rlsio})) holds for
\emph{all} elements in  ${\cal B}^2_{e,\mu}$ (resp., ${\cal B}^2_{o,\mu}$)
due to an argument based on Corollary \ref{maincor}.
Again, see \cite{RLSI} for more details.
Similar reasoning justifies the subsequent results which, at first glance,
appear to hold only in a certain dense subspace of the relevant
Hilbert space, but actually hold in all of that Hilbert space.

The inequality (\ref{rlsie}) should be compared with
Theorem~5.1 in \cite{AS},
which says in our notation that
$$
             E_{e,\mu} (f_e) \le c S_{ L^2 ( \nu_{e,\mu} ) } ( f_e ) 
         + \kappa_e(c,\mu) || f_e ||^2_{ L^2 ( \nu_{e,\mu} ) }. 
$$
We have shown in \cite{AS} that for each $c>1$ we can take
$$
   \kappa_e(c,\mu) = c \log \int_{ {\bf C} } d\nu_{e,\mu} (z) \, e^{|z|^2/c} 
< \infty.
$$
So we have proved the same type of reverse log-Sobolev inequality in ${\cal B}^2_{e,\mu}$ 
though with a possibly different coefficient for the norm term. 
Similarly, our result (\ref{rlsio}) in the odd case corresponds to Theorem~5.2 in \cite{AS} with the same caveats. 
The method of \cite{AS} is based on the Young inequality and is due to Gross. 
(See \cite{GGS} and \cite{SOME}.)
One advantage of the results in \cite{AS} is that 
formulas are produced for the coefficients of the norm terms. 
Our analysis here is incomplete in that regard.
It remains an open problem to identify the optimal constants of the norm terms.
They may even be equal to zero as far as we currently know. 

   Finally, Corollary~\ref{corevenodd} in the particular case that
$p_e=p_o$, $q_e=q_o$ and $a_e=a_o$ with $p_e^\prime q_e / 4 -1 < a_e < 0$
reduces to
\begin{eqnarray*} 
E_\mu(f) &\le& c \left(
S_{ L^2 \left( \nu_{e,\mu} \right) } \left( f_e \right) 
+ 
S_{ L^2 \left( \nu_{o,\mu} \right) } \left( f_o \right) 
\right) 
\nonumber \\
&+& \tau_e(c,\mu) || f_e ||^2_{ L^2 \left( \nu_{e,\mu} \right) } 
+
\tau_o(c,\mu) || f_o ||^2_{ L^2 \left( \nu_{o,\mu} \right) }, 
\nonumber
\end{eqnarray*}
for any $c > 1$, 
using the definition of $E_\mu(f)$ in equation (\ref{defemu}). 
This is the first inequality in Theorem~5.3 in \cite{AS}, modulo the coefficient of the norm term.
By taking $\tau(c,\mu):= \max ( \tau_e(c,\mu), \tau_o(c,\mu) )$, we get the next result.
\begin{theorem}
\label{thm45}
(Reverse log-Sobolev inequality in ${ {\cal B}^2_\mu }$ for $\mu$-deformed energy.) 
For every $f = f_e + f_o \in { {\cal B}^2_\mu }$ we have that
\begin{equation} 
\label{star1}
E_\mu(f) \le c \left\{
S_{ L^2 \left( \nu_{e,\mu} \right) } \left( f_e \right) 
+ 
S_{ L^2 \left( \nu_{o,\mu} \right) } \left( f_o \right) 
\right\}
+ \tau(c,\mu) || f ||^2_{ {\cal B}^2_\mu }
\end{equation}
for any $c>1$, where $\tau(c,\mu)$ is a finite constant. 
\end{theorem} 
This is the second inequality in Theorem~5.3 of \cite{AS}, 
again modulo the coefficient of the norm term.
Note that this does not appear to be a reverse
log-Sobolev inequality in the sense of Definition~\ref{def14}
given that the expression in brackets on the right side of (\ref{star1}) may
not be immediately seen to be a Shannon entropy.
In fact, it is \emph{not} a Shannon entropy of $f \in { {\cal B}^2_\mu }$,
since ${ {\cal B}^2_\mu }$ is not defined as a subspace of an $L^2$ space.
And we stated just this in \cite{AS}, but it turns out 
that there is another way of viewing this.
Note that the isometry $f \to (f_e , f_o) $ maps
$${ {\cal B}^2_\mu } \to
L^2({\bf C}, \nu_{e,\mu}) \oplus L^2({\bf C}, \nu_{o,\mu})
\cong
L^2({\bf C} \times {\bf Z}_2, \nu_\mu)
$$
as we remarked in Section~3
and so this canonically identifies ${ {\cal B}^2_\mu }$ with a closed
subspace of $L^2({\bf C} \times {\bf Z}_2, \nu_\mu)$, which is an
$L^2$ space.
We use this fact in the next definition.

\begin{definition}
\label{defmudeent}
(See \cite{AS}.)
For $f = f_e + f_o \in {\cal B}^2_\mu$ we define its 
\emph{$\mu$-deformed entropy} by
$$ 
   S_{\mu}(f) := S_{L^2({\bf C} \times {\bf Z}_2, \nu_\mu)}(f_e,f_o).
$$ 
\end{definition}
Then we immediately calculate
$S_{\mu}(f) = S_{ L^2(\nu_{e,\mu})} (f_e)  + S_{ L^2(\nu_{o,\mu})} (f_o)$,
which agrees with the definition in \cite{AS}.
Now this allows us to write (\ref{star1}) as follows:
$$
E_\mu(f) \le c S_{\mu}(f) 
+ \tau(c,\mu) || f ||^2_{ {\cal B}^2_\mu }. 
$$

   In summary, we have another method for proving the reverse log-Sobolev 
inequalities in \cite{AS}.
However, the coefficients of the norm terms that we obtain here are most likely different. 
(They \emph{are} different in the case $\mu=0$. See \cite{SOME}.) 

   An important point is that the reproducing kernel method also produces direct 
log-Sobolev inequalities in ${\cal B}^2_{e,\mu}$ and in ${\cal B}^2_{o,\mu}$, and
these are new results. 
So, we have the next result, which is a restatement of
Case~1 of Theorem~\ref{thm43}
\begin{theorem}
\label{thm46}
(Log-Sobolev inequalities in ${\cal B}^2_{e,\mu}$ and ${\cal B}^2_{o,\mu}$ for
the $\mu$-deformed energy.) 
For all $f_e \in {\cal B}^2_{e,\mu}$ we have 
$$
       a_e S_{ L^2(\nu_{e,\mu})} (f_e) \le 
       b_e E_{e,\mu} (f_e)  
       + c_e ||f_e||^2_{ L^2(\nu_{e,\mu})},
$$ 
where $a_e > 0$, $b_e > 0$ and $c_e \ge 0$ are finite constants. \\
For all $f_o \in {\cal B}^2_{o,\mu}$ we have 
$$
       a_o S_{ L^2(\nu_{o,\mu})} (f_o) \le 
       b_o E_{o,\mu} (f_o)  
       + c_o ||f_o||^2_{ L^2(\nu_{o,\mu})},
$$ 
where $a_o > 0$, $b_o > 0$ and $c_o \ge 0$ are finite constants. 
\end{theorem}
Of course, one can divide both sides in the previous inequalities by the coefficient
of the entropy term without changing the sense of the inequality. 
Then one would try to find the optimal constant for the norm term, given a fixed value 
for the coefficient of the energy term. 

   Next by summing these two direct log-Sobolev inequalities,
we obtain an energy-entropy inequality in
${\cal B}^2_\mu$ with two entropy terms of the form:
\begin{eqnarray*} 
       a_e S_{ L^2(\nu_{e,\mu})} (f_e)  + a_o S_{ L^2(\nu_{e,\mu})} (f_o)  &\le& 
       b_e E_{e,\mu} (f_e) + b_o E_{o,\mu} (f_o) \\ 
       &+& c_e ||f_e||^2_{ L^2(\nu_{e,\mu})} + c_o ||f_o||^2_{ L^2(\nu_{o,\mu})}, 
\end{eqnarray*} 
where $f = f_e + f_o \in {\cal B}^2_\mu$. 
By taking $a := \min ( a_e, a_o)$, $b:= \max (b_e, b_o)$ and $c:= \max (c_e, c_o)$, 
we get for all $f = f_e + f_o \in {\cal B}^2_\mu$ that
$$ 
       a \left\{ S_{ L^2(\nu_{e,\mu})} (f_e)  + S_{ L^2(\nu_{e,\mu})} (f_o) \right\} \le 
       b E_{\mu} (f) + c ||f||^2_{ {\cal B}^2_\mu }. 
$$
We can apply Definition~\ref{defmudeent} to the term in brackets 
on the left hand side and get the next result.
\begin{theorem}
(Log-Sobolev inequality for ${\cal B}^2_\mu$ for the $\mu$-deformed energy.) 
For all $f \in {\cal B}^2_\mu$ we have that 
$$ 
       a S_{\mu}(f)  \le 
       b E_{\mu} (f) + c ||f||^2_{ {\cal B}^2_\mu }, 
$$ 
where $a>0$, $b>0$ and $c \ge 0$ are finite constants. 
\end{theorem}

   As a closing comment to this section, we note that some other rather 
strange looking inequalities can be obtained from these results.  For 
example, we can add a direct log-Sobolev inequality for ${\cal B}^2_{e,\mu}$ 
with a reverse log-Sobolev inequality for ${\cal B}^2_{e,\mu}$. 
(Similarly, we can do this for ${\cal B}^2_{o,\mu}$.) 
We can also add a direct log-Sobolev inequality for ${\cal B}^2_{e,\mu}$ 
with a reverse log-Sobolev inequality for ${\cal B}^2_{o,\mu}$ and, 
vice versa, a direct log-Sobolev inequality for ${\cal B}^2_{o,\mu}$ 
with a reverse log-Sobolev inequality for ${\cal B}^2_{e,\mu}$. 
Of course, none of these inequalities is more fundamental than their antecedents, 
and they seem to be mere curiosities as far as we can tell. 

\section{Dirichlet and $\mu$-deformed energies}

The $\mu$-deformed energies introduced by us in \cite{AS} can be related to a Dirichlet 
form energy in ${\cal B}^2_\mu$. 
So we proceed to a discussion that will lead us to a definition of this latter concept. 

We first note that one can introduce creation and annihilation operators,  
$A^*_\mu$ and $A_\mu$ respectively, which act in ${\cal B}^2_\mu$. 
In terms of the standard orthonormal basis $\{ \Psi^\mu_n \}_{n \ge 0}$ of ${\cal B}^2_\mu$,
where $ \Psi^\mu_n (z) = z^n / ( \gamma_\mu (n) )^{1/2}$ (see \cite{MA}), the definitions are: 
\begin{gather}
\label{defa}
     A_\mu \Psi^\mu_n := \left( \frac{\gamma_\mu(n)}{\gamma_\mu(n-1)} \right)^{1/2} \Psi^\mu_{n-1} \\
\label{defastar}   
 A^*_\mu \Psi^\mu_n := \left( \frac{\gamma_\mu(n+1)}{\gamma_\mu(n)} \right)^{1/2} \Psi^\mu_{n+1} 
\end{gather}
for every integer $n \ge 0$, where $\Psi^\mu_{-1} \equiv 0$.
Then, one can extend the definitions (\ref{defa}) and (\ref{defastar}) linearly to 
the dense subspace  ${\cal D}^2_\mu$ of  ${\cal B}^2_\mu$, where ${\cal D}^2_\mu$
is defined to be the set of all finite linear combinations of the $\Psi^\mu_n$.  
While we have given these definitions explicitly in \cite{AS}, one can find them discussed in  
a quite general situation in Section~5 of Rosenblum's article \cite{RO} and, in a form isomorphic 
to that given here, in formulas (3.7.1) and (3.7.2) of \cite{RO}. 
Moreover, it can be easily checked that 
\begin{gather} 
\label{defd} 
A_\mu f (z) = D_\mu f (z) := f^\prime (z) + \frac{\mu}{z} \left( f \left( z \right) - f \left( -z \right) \right) \\
\label{defm}
A^*_\mu f (z) = (M_\mu f) (z) := z f(z)
\end{gather}
for all $ f \in {\cal D}^2_\mu$ and all $ z \in {\bf C}$. 
Here $f^\prime (z)$ is the complex derivative of $f(z)$. 
(We thank C.~Pita for bringing formula (\ref{defd}) to our attention.) 
Of course, the formulas (\ref{defd}) and (\ref{defm}) can be used to define  
$D_\mu$ and $M_\mu$, and hence $A_\mu$ and $A^*_\mu$ as well, on much 
larger spaces than ${\cal D}^2_\mu$. 
For example, we will use these formulas for definitions on ${\cal B}^2_\mu$ 
with the warning that the range will not then be a subspace of ${\cal B}^2_\mu$. 
We also use these formulas for definitions on ${\cal H}({\bf C})$, the space of
all holomorphic functions on ${\bf C}$, which is a domain invariant under the 
actions of $D_\mu$ and $M_\mu$. 
(Note that the singularity  at $z=0$ in the second term of (\ref{defd}) is removable since $f$ is holomorphic.) 
The operators $D_\mu$ and $M_\mu$ already appear in \cite{RO}, p.~373. 
Moreover, $D_\mu$ is well known to be a special case of a Dunkl operator. 
(See \cite{MR} and references therein.) 
From equations (\ref{defd}) and (\ref{defm}) one sees immediately that 
\begin{equation}
\label{ccrmu}
              [ A_\mu, A^*_\mu ] = I + 2 \mu J
\end{equation} 
on ${\cal H}({\bf C})$. 
Of course, $ [ A_\mu, A^*_\mu ] = A_\mu A^*_\mu - A^*_\mu A_\mu$ 
is the usual commutator of the two operators $A_\mu$ and $A^*_\mu$,
$I$ is the identity operator, and $J$ is the parity operator as introduced earlier. 
The commutation relation (\ref{ccrmu}), which differs from the canonical 
commutation relation in the second term on the right, was essentially 
introduced by Wigner in \cite{WI} in order to answer negatively the 
question whether the standard quantum mechanical equations of motion 
determine the canonical commutation relations. 
Actually, Wigner presented
a commutation relation for $\mu$-deformed position and momentum operators 
($Q_\mu$ and $P_\mu$) that 
is equivalent to (\ref{ccrmu}). 
The article \cite{WI} by Wigner is the starting point of all further research concerning
operators like $Q_\mu$, $P_\mu$, $A_\mu$ and $A^*_\mu$ 
and the spaces on which they act.

     Up to this point in the discussion, $A_\mu$ and $A^*_\mu$ are two operators, 
each with its own definition.
More than anything else, the notation indicates a wish that $A_\mu$ and $A^*_\mu$ 
should be adjoints of each other. 
But to define adjoints, one needs an inner product, and such a 
structure is not available in ${\cal H} ( {\bf C} )$. 
However, we can realize $A_\mu$ and $A^*_\mu$ as densely defined, 
closed unbounded operators in the Hilbert space ${\cal B}^2_\mu$. 
Then we do have the adjointness relation 
$$
\left\langle A^*_\mu f,  g \right\rangle_{ {\cal B}^2_\mu } = 
\left\langle f,  A_\mu g \right\rangle_{ {\cal B}^2_\mu }
$$
for all $f$ in the domain of $A^*_\mu$ and 
for all $g$ in the domain of $A_\mu$. 
As discussed further in \cite{SOGTMP}, this relation can be taken as the 
motivation for the definition of the inner product for ${\cal B}^2_\mu$. 

   The \emph{$\mu$-deformed number operator}  (see \cite{AS}) is defined by 
$$
    N_\mu := A^*_\mu A_\mu = M_\mu D_\mu,
$$
and its associated quadratic form is then 
\begin{equation} 
\label{defqform}
\left\langle f , N_\mu f \right\rangle_{ {\cal B}^2_\mu } 
= \left\langle f, A^*_\mu A_\mu f \right\rangle_{ {\cal B}^2_\mu } 
= \left\langle A_\mu f,  A_\mu f \right\rangle_{ {\cal B}^2_\mu } 
= || D_\mu f ||^2_{ {\cal B}^2_\mu }.
\end{equation}
This last expression justifies our calling this 
a \emph{Dirichlet form}. 

    While the left side of (\ref{defqform}) has a natural domain given by the domain of $N_\mu$, 
the right side has a natural domain given by the domain of $D_\mu$, which is strictly larger. 
Specifically we have
\begin{gather*}
                {\rm Domain} (N_\mu) =  \{ f \in  {\cal B}^2_\mu ~:~ N_\mu f \in  {\cal B}^2_\mu \}, \\
                {\rm Domain} (D_\mu) =  \{ f \in  {\cal B}^2_\mu ~:~ D_\mu f \in  {\cal B}^2_\mu \}. 
\end{gather*}

\begin{definition} 
The \emph{Dirichlet form energy}  
(or the \emph{Dirichlet energy}) is defined 
as $|| D_\mu f ||^2_{ {\cal B}^2_\mu }$
for all $f$ in ${\rm Domain} (D_\mu)$ and as $+\infty$ otherwise. 
\end{definition}
We avoid the standard convention of writing 
$\left\langle f , N_\mu f \right\rangle_{ {\cal B}^2_\mu }$ for the Dirichlet energy. 
In fact, the operator $N_\mu$ does not enter the discussion here in any essential
way, and we will not make any further explicit reference to it.

   Note that we can use the commutation relation (\ref{ccrmu}) to obtain, at least formally,
\begin{gather*}
|| D_\mu f ||^2_{ {\cal B}^2_\mu } = \left\langle A_\mu f , A_\mu f \right\rangle_{ {\cal B}^2_\mu } 
= \left\langle f, A^*_\mu A_\mu f \right\rangle_{ {\cal B}^2_\mu } \nonumber \\
= \left\langle f, ( A_\mu A^*_\mu - I - 2\mu J) f \right\rangle_{ {\cal B}^2_\mu } \nonumber \\
= || A^*_\mu f ||^2_{ {\cal B}^2_\mu } - || f ||^2_{ {\cal B}^2_\mu } - 2 \mu \left\langle f, Jf \right\rangle_{ {\cal B}^2_\mu }. 
\end{gather*} 
To make this rigorous, 
we will use the next result, whose proof is elementary. 
(See \cite{BA} for a proof in the case $\mu=0$.) 
\begin{prop} 
Suppose $g(z) = \sum_{k=0}^\infty b_k z^k$ for $b_k \in {\bf C}$ is an entire function, that is, 
it is holomorphic for all $z \in {\bf C}$. 
Then, 
\begin{equation} 
\label{gequals} 
        || g ||^2_{ {\cal B}^2_\mu } = \sum_{k=0}^\infty | b_k |^2 \gamma_\mu (k), 
\end{equation} 
where both sides are defined to be elements in $[0,\infty]$. 
In particular, $g \in {\cal B}^2_\mu $ if and only if the series on the right hand side 
of (\ref{gequals}) is convergent. 
\end{prop} 

     We now prove the result which we derived formally above. 

\begin{prop}
For all $f \in {\cal B}^2_{\mu}$ we have that 
\begin{equation}
\label{formalid}
|| D_\mu f ||^2_{ {\cal B}^2_\mu } = 
|| A^*_\mu f ||^2_{ {\cal B}^2_\mu } - || f ||^2_{ {\cal B}^2_\mu } - 2 \mu \left\langle f, Jf \right\rangle_{ {\cal B}^2_\mu }. 
\end{equation}
In particular, $|| D_\mu f ||_{ {\cal B}^2_\mu } < \infty$ if and only if 
$|| A^*_\mu f ||_{ {\cal B}^2_\mu } < \infty$. 
\end{prop}
{\bf Proof:} 
First we write $ f(z) = \sum_{k=0}^\infty a_{k} z^{k}$, and we then calculate that 
\begin{gather*}
        D_\mu f(z) = \sum_{k=0}^\infty a_{k} \left(k + 2\mu \chi_o \left( k \right) \right) z^{k-1}, \\
        A^*_\mu f (z) = \sum_{k=0}^\infty a_{k} z^{k+1}, \\ 
                       Jf(z) = \sum_{k=0}^\infty (-1)^k a_{k} z^{k}, 
\end{gather*} 
where $\chi_o$ is the characteristic function of the odd integers. 
It then  follows that
\begin{gather*}
    || D_\mu f ||^2_{ {\cal B}^2_\mu } = \sum_{k=0}^\infty |a_{k}|^2 (k + 2\mu \chi_o (k) )^2 \gamma_\mu(k-1), \\ 
    || A^*_\mu f ||^2_{ {\cal B}^2_\mu } = \sum_{k=0}^\infty |a_{k}|^2 \gamma_\mu(k+1), \\
    || f ||^2_{ {\cal B}^2_\mu } = \sum_{k=0}^\infty |a_{k}|^2 \gamma_\mu(k),  \\
    \left\langle f, Jf \right\rangle_{ {\cal B}^2_\mu } = \sum_{k=0}^\infty  (-1)^k |a_{k}|^2 \gamma_\mu(k). 
\end{gather*} 
Here we use the convention that $\gamma_\mu (-1) = 0$. 
So (\ref{formalid}) is a direct consequence of 
$$ 
(k + 2\mu \chi_o (k) )^2 \gamma_\mu(k-1) = \gamma_\mu(k+1) - \gamma_\mu(k) - (2 \mu) (-1)^k \gamma_\mu(k)
$$ 
for all integers $k \ge 0$, 
which in turn follows from the definition (\ref{defgamma}) of the $\mu$-deformed factorial $\gamma_\mu$. 
Note that we have proved (\ref{formalid}) for \emph{all} $ f \in {\cal B}^2_{\mu}$ in the sense 
that one side is finite if and only if the other side is finite.  
Since the last two terms on the right hand side of  (\ref{formalid}) are finite for all
$ f \in {\cal B}^2_{\mu}$,
the last assertion of the theorem follows directly. QED.

The previous two propositions also appear in \cite{SISOL}.

Notice that 
\begin{equation} 
\label{astarsq} 
|| A^*_\mu f ||^2_{ {\cal B}^2_\mu } = || M_\mu f ||^2_{ {\cal B}^2_\mu } 
= \int_{ {\bf C} } d\nu_{e,\mu} (z) |z|^2 |f_o (z)|^2 +  \int_{ {\bf C} } d\nu_{o,\mu} (z) |z|^2 |f_e (z)|^2, 
\end{equation} 
since $ (z f(z))_e = z f_o(z)$ and $ (z f(z))_o = z f_e(z)$. 
While the last two integrals in (\ref{astarsq}) are reminiscent of the $\mu$-deformed energies, 
$E_{e,\mu}(f_e)$ and $E_{o,\mu}(f_o)$, they are in fact \emph{new} quantities. 
One way to think of this is that the integrals in (\ref{astarsq}) are ``mixed'' 
in terms of parity in the sense that 
the expression involving $f_o$ in the first integral 
is integrated with respect to $d\nu_{e,\mu}$ and, vice versa, 
the expression involving $f_e$ in the second integral is integrated with respect to $d\nu_{o,\mu}$. 
However, in $E_{e,\mu}(f_e)$ an even function $f_e$ is integrated with respect to
$d\nu_{e,\mu}$,
and in $E_{o,\mu}(f_o)$ an odd function $f_o$ is integrated with respect to
$d\nu_{o,\mu}$.

   The question now is how to relate the $\mu$-deformed energies
to these new quantities 
on the right side on (\ref{astarsq}), and hence to the Dirichlet energy.
First off, consider the case $\mu > 0$. 
The inequality $ \nu_{e,\mu}(z) < \nu_{o,\mu}(z)$ 
of densities for $ 0 \ne z \in {\bf C}$ 
given in (\ref{eleo}) allows us to write 
for  $0 \ne f_e \in {\cal B}^2_{e,\mu}$ that
\begin{equation} 
\label{emue} 
E_{e,\mu} (f_e) = \int_{ {\bf C} } d\nu_{e,\mu} (z) |z|^2 | f_e (z) |^2 
< \int_{ {\bf C} } d\nu_{o,\mu} (z) |z|^2 | f_e (z) |^2 .
\end{equation} 
Similarly, for  $0 \ne f_o \in {\cal B}^2_{o,\mu}$ we have that
\begin{equation} 
\label{omuo}
\int_{ {\bf C} } d\nu_{e,\mu} (z) |z|^2 | f_o (z) |^2 
<  \int_{ {\bf C} } d\nu_{o,\mu} (z) |z|^2 | f_o (z) |^2 = E_{o,\mu} (f_o).
\end{equation} 
However, for $\mu > 0$, we do {\em not} have an inequality $\nu_{o,\mu}(z) \le C \nu_{e,\mu}(z)$ 
as we can see from the asymptotic behavior near zero of each side. 
Nonetheless, we claim that reverse inequalities corresponding to (\ref{emue}) and (\ref{omuo}) 
can be proved. 
The complete result for all the possible cases for $\mu$ is as follows.

\begin{theorem}
\label{thm51}
For every $\mu > 0$ there exists positive constants $C_{e,\mu} > 1$ and $C_{o,\mu} < 1$  such that 
\begin{equation}
\label{firstresult}
      E_{e,\mu}(f_e) < \int_{ {\bf C} } d\nu_{o,\mu} (z) |z|^2 |f_e(z)|^2 \le C_{e,\mu} E_{e,\mu}(f_e) 
\end{equation}
for all $0 \ne f_e \in {\cal B}^2_{e,\mu}$ and 
$$ 
      C_{o,\mu} E_{o,\mu}(f_o) \le \int_{ {\bf C} } d\nu_{e,\mu} (z) |z|^2 |f_o(z)|^2 < E_{o,\mu}(f_o) 
$$ 
for all $0 \ne f_o \in {\cal B}^2_{o,\mu}$. \\
For the case $\mu =0$, we have that 
$$
      E_{e,0}(f_e) = \int_{ {\bf C} } d\nu_{o,0} (z) |z|^2 |f_e(z)|^2
$$
and
$$
      E_{o,0}(f_o) = \int_{ {\bf C} } d\nu_{e,0} (z) |z|^2 |f_o(z)|^2. 
$$
Finally, for the case $-1/2 < \mu <0$ we have 
$$ 
      E_{e,\mu}(f_e) > \int_{ {\bf C} } d\nu_{o,\mu} (z) |z|^2 |f_e(z)|^2 \ge C_{e,\mu} E_{e,\mu}(f_e) 
$$ 
for all $0 \ne f_e \in {\cal B}^2_{e,\mu}$ and 
$$ 
      C_{o,\mu} E_{o,\mu}(f_o) \ge \int_{ {\bf C} } d\nu_{e,\mu} (z) |z|^2 |f_o(z)|^2 > E_{o,\mu}(f_o) 
$$ 
for all $0 \ne f_o \in {\cal B}^2_{o,\mu}$, where $0 < C_{e,\mu} < 1$ and $C_{o,\mu} > 1$. 
\end{theorem} 
\vskip .3cm 
\noindent
{\bf Proof:} 
Suppose that $f_e \in {\cal B}^2_{e,\mu}$. 
We claim that $E_{e,\mu} (f_e) < \infty$ 
if and only if  $ \int_{ {\bf C} } d\nu_{o,\mu} (z) |z|^2 | f_e (z) |^2 < \infty$. 
Actually, $E_{e,\mu} (f_e) < \infty$ if and only if 
\begin{equation} 
\label{nearinfinity}
 \int_{\bf C} dx dy \, |z|^{2\mu+3} \exp ( -|z|^2 ) \, |f_e(z)|^2 < \infty
\end{equation}
by the asymptotic behavior of the Macdonald function $K_{\mu - 1/2}$ near infinity.  
The point here is that $f_e$ has no local singularities, being holomorphic, and so only
its behavior near infinity matters for the convergence of the integral that defines $E_{e,\mu} (f_e) $. 
But $ \int_{ {\bf C} } d\nu_{o,\mu} (z) |z|^2 | f_e (z) |^2 < \infty$ if and only if 
(\ref{nearinfinity}) holds, 
since again only the
asymptotic behavior near infinity matters, and the behavior of $K_{\mu + 1/2}$ to first order 
near infinity is the same as that of $K_{\mu - 1/2}$ near infinity. 
This establishes the claim. 
(Actually, in this part of the proof only the continuity of $f_e$ plays a role.) 
The expressions 
$$ 
\left(|| f_e ||^2 + E_{e,\mu} (f_e) \right)^{1/2} 
$$ 
and
$$
\left( || f_e ||^2 + \int_{ {\bf C} } d\nu_{o,\mu} (z) |z|^2 | f_e (z) |^2 \right)^{1/2}
$$ define 
Hilbert norms in ${\cal B}^2_{e,\mu}$, and
the result of the previous paragraph says that they define the same finite norm subspace, say ${\cal F}$, 
of ${\cal B}^2_{e,\mu}$. 
Moreover, this subspace ${\cal F}$ is closed in the corresponding entire $L^2$ space with respect to 
either one of these norms, and so ${\cal F}$ is a 
Hilbert space with respect to either one of these norms.
(It is at this point that the holomorphicity of the functions is used in a standard argument already seen in Proposition \ref{prop11}.)
We now consider the case $\mu > 0$. 
But then the open mapping theorem (See \cite{RS}, p. 82.) together with 
the first inequality in (\ref{firstresult}),
which we proved just before stating this theorem, 
implies the second inequality in (\ref{firstresult}) for all $f \in {\cal F}$. 
But (\ref{firstresult}) is trivially true for all $f \in {\cal B}^2_{e,\mu} \setminus {\cal F}$, 
since all three expressions are then equal to $+\infty$. 

   The case when $-1/2 < \mu < 0$ follows by similar arguments.
Finally, the case $\mu = 0$ follows from the fact that $d\nu_{e,0} = d\nu_{o,0}$, something that we 
have already noted. QED.
\vskip .2cm \noindent
{\bf Remark:}  It would be desirable to give a constructive proof of this theorem 
for the case $\mu \ne 0$ with explicit formulas for 
$C_{e,\mu}$ and $C_{o,\mu}$. 
It also remains an open problem to identify the optimal values for the constants 
$C_{e,\mu}$ and $C_{o,\mu}$ when $\mu \ne 0$. 

      Though we will not use the next result in the form stated, we feel it 
is worthwhile to include it here since it is the idea behind the remaining 
results in this section.  
It is an immediate consequence of (\ref{formalid}), (\ref{astarsq}) and Theorem~\ref{thm51}.
\begin{corollary}
We have the following equivalences of $\mu$-deformed and Dirichlet energies:
\begin{enumerate}
\item
For all $g \in {\cal B}^2_{e,\mu}$ we have that the $\mu$-deformed energy 
$E_{e,\mu} (g)$ is finite if and only if the Dirichlet energy
$|| D_\mu g ||^2_{  {\cal B}^2_\mu }$ is finite.
\item
For all $h \in {\cal B}^2_{o,\mu}$ we have that the $\mu$-deformed energy 
$E_{o,\mu} (h)$ is finite if and only if the Dirichlet energy
$|| D_\mu h ||^2_{  {\cal B}^2_\mu }$ is finite.
\item
For all $f \in {\cal B}^2_\mu$ we have that the $\mu$-deformed energy 
$E_{\mu} (f)$ is finite if and only if the Dirichlet energy
$|| D_\mu f ||^2_{  {\cal B}^2_\mu }$ is finite.
\end{enumerate}
\end{corollary}

   We can now put together the results of Section~4 and Theorem~\ref{thm51} to
get direct and reverse inequalities for the Dirichlet energy
$|| D_\mu f ||^2_{  {\cal B}^2_\mu }$ and Shannon entropy. 
We continue using the notation from Section~4 and Theorem~\ref{thm51}.
We only state the case $\mu \ge 0$.
The case $-1/2 < \mu <0$ is quite similar.

\begin{theorem}
\label{thm52}
(Reverse log-Sobolev inequalities in ${\cal B}^2_{e,\mu}$ and ${\cal B}^2_{o,\mu}$ 
for Dirichlet energy.) 
Suppose that $\mu \ge 0$ and that $c >1$. 
For every $f_e \in {\cal B}^2_{e,\mu}$ we have that 
\begin{equation*} 
              || D_\mu f_e ||^2_{  {\cal B}^2_\mu }
              \le c C_{e,\mu} S_{ L^2 ( \nu_{e,\mu} ) } ( f_e ) 
         + ( C_{e,\mu} \tau_e(c) - (1+2\mu) ) || f_e ||^2_{ L^2 ( \nu_{e,\mu} ) }.
\end{equation*} 
For every $f_o \in {\cal B}^2_{o,\mu}$ we have that 
\begin{equation*} 
                || D_\mu f_o ||^2_{  {\cal B}^2_\mu }
              \le c S_{ L^2 ( \nu_{o,\mu} ) } ( f_o ) 
         + ( \tau_o(c) - (1 - 2\mu) ) || f_o ||^2_{ L^2 ( \nu_{o,\mu} ) }.
\end{equation*}
For every $f = f_e + f_o \in { {\cal B}^2_\mu }$ we have that 
\begin{gather*} 
   || D_\mu f ||^2_{  {\cal B}^2_\mu } \le 
c C_{e,\mu} S_{ L^2 ( \nu_{e,\mu} ) } ( f_e ) + c S_{ L^2 ( \nu_{o,\mu} ) } ( f_o ) \\ 
+ ( C_{e,\mu} \tau_e(c) - (1+2\mu) ) || f_e ||^2_{ L^2 ( \nu_{e,\mu} ) } 
 + ( \tau_o(c) - (1 - 2\mu) ) || f_o ||^2_{ L^2 ( \nu_{o,\mu} ) }. 
\end{gather*}
\end{theorem} 
{\bf Proof:} The first two inequalities follow immediately from 
Theorems~\ref{thm45} and \ref{thm51} 
as well as the identities (\ref{formalid}) and (\ref{astarsq}).
The last inequality is the sum of the previous two inequalities. 
It can be simplified a bit by estimating the sum of the norm terms.
QED.

\begin{theorem}
\label{thm53}
(Log-Sobolev inequalities in ${\cal B}^2_{e,\mu}$ and ${\cal B}^2_{o,\mu}$ 
for Dirichlet energy.) 
Suppose that $\mu \ge 0$. 
Then there are real constants $a_e > 0$, $b_e > 0$ and $c_e \ge 0$ such that
for all $f_e \in {\cal B}^2_{e,\mu}$ we have
$$
       a_e S_{ L^2(\nu_{e,\mu})} (f_e) \le
       b_e || D_\mu f_e ||^2_{  {\cal B}^2_\mu } 
       + (b_e(1+2\mu) + c_e) ||f_e||^2_{ L^2(\nu_{e,\mu})}.
$$
Also there are real constants $a_o > 0$, $b_o > 0$ and $c_o \ge 0$ such that
for all $f_o \in {\cal B}^2_{o,\mu}$ we have
$$
       a_o S_{ L^2(\nu_{o,\mu})} (f_o) \le 
          b_o C_{o,\mu}^{-1} || D_\mu f_o ||^2_{  {\cal B}^2_\mu } 
       + ( b_o C_{o,\mu}^{-1} (1-2\mu) + c_o) ||f_o||^2_{ L^2(\nu_{o,\mu})}. 
$$ 
Finally, for every $f = f_e + f_o \in { {\cal B}^2_\mu }$ we have that
\begin{gather*} 
    a_e S_{ L^2(\nu_{e,\mu})} (f_e) + a_o S_{ L^2(\nu_{o,\mu})} (f_o) \le
    b_e || D_\mu f_e ||^2_{  {\cal B}^2_\mu } +   b_o C_{o,\mu}^{-1} || 
    D_\mu f_o ||^2_{  {\cal B}^2_\mu } \\
    + (b_e(1+2\mu) + c_e) ||f_e||^2_{ L^2(\nu_{e,\mu})}  +
    ( b_o C_{o,\mu}^{-1} (1-2\mu) + c_o) ||f_o||^2_{ L^2(\nu_{o,\mu})}.
\end{gather*}
\end{theorem}
{\bf Proof:} The first two inequalities follow immediately from
Theorems~\ref{thm46} and \ref{thm51}
as well as the identities (\ref{formalid}) and (\ref{astarsq}).
The last inequality is the sum of the previous two inequalities. 
It can also be simplified in form by using appropriate trivial estimates.
QED.

It seems reasonable to conjecture that the inequalities in Theorem~\ref{thm53}
hold without the norm
term, since this is known to be true in the case $\mu = 0$.
However, the situation is not as clear for Theorem~\ref{thm52}.
It remains an open problem to
determine the optimal coefficient of the norm term for each of these inequalities
in Theorems~\ref{thm52} and \ref{thm53},
given that the other coefficients are fixed. 

  Just as in the previous section, we obtain the next immediate but important
consequence.
\begin{corollary} 
We have these equivalences of entropies and Dirichlet energies:
\begin{enumerate}
\item
For all $g \in {\cal B}^2_{e,\mu}$ we have that the Shannon entropy 
$S_{ L^2(\nu_{e,\mu})} (g)$ is finite if and only if the Dirichlet energy  
$|| D_\mu g ||^2_{  {\cal B}^2_\mu }$ is finite. 
\item
For all $h \in {\cal B}^2_{o,\mu}$ we have that the Shannon entropy 
$S_{ L^2(\nu_{o,\mu})} (h)$ is finite if and only if the Dirichlet energy  
$|| D_\mu h ||^2_{  {\cal B}^2_\mu }$ is finite.   
\item
For all $f \in {\cal B}^2_\mu$ we have that the $\mu$-deformed entropy 
$S_{\mu}(f)$ is finite if and only if the Dirichlet energy  
$|| D_\mu f ||^2_{  {\cal B}^2_\mu }$ is finite. 
\end{enumerate}
\end{corollary}

In \cite{PS1} another quadratic form, called the \emph{dilation energy},
is introduced in the $\mu$-deformed Segal-Bargmann space.
It is shown there that this dilation energy is comparable to the
$\mu$-deformed energy.
So it is straightforward to obtain results analogous to those in this
section with the dilation energy replacing the $\mu$-deformed energy.
The details are left to the interested reader.
Actually, the log-Sobolev inequality proved in Theorem~6.3
of \cite{PS1} can be used to prove a log-Sobolev inequality in
the Segal-Bargmann space, though
those authors did not state this explicitly.
Nor did we realize this until we concluded this article.
It turns out that the log-Sobolev inequality proved in \cite{PS1}
has a very different flavor to it, since in general it
relates entropies in two different spaces to each other
much in the manner of a Hirschman inequality.

\section{Concluding Remarks} 
Besides the problem of determining the best constants for all of the inequalities
proved here, another open problem is to establish a hypercontractivity result 
for this scale of $\mu$-deformed Segal-Bargmann spaces. 
Note that in \cite{AS} we have shown reverse hypercontractivity 
in this scale of spaces. 

We can consider formulating this theory in terms of holomorphic functions
defined on ${\bf C}^n$ instead of on ${\bf C}$. 
This can be done where one replaces the Coxeter group
${\bf Z}_2 = \{ I, J \}$ used here 
with the Coxeter group $({\bf Z}_2)^n$ generated by the reflections $J_k$ in ${\bf C}^n$ 
given by $J_k ( z_1, \dots ,  z_k, \dots , z_n) := ( z_1, \dots , -z_k, \dots , z_n)$ for
$k = 1, \dots , n$. 
We thank C.~Pita for telling us about this formulation, which is also discussed in \cite{SBSBO}.
However, the resulting theory is in some sense trivial in that everything factorizes 
as an $n$-fold product of the structures discussed here. 
It may be the case that with other choices of Coxeter group the theory in 
dimension $n$ could be non-trivial. 
Refer to \cite{SBSBO} for more details. 
Of course, there is also the possibility of doing this sort of theory in infinite dimension.

   Finally, there is a ``configuration'' space $L^2( {\bf R}, |x|^{2\mu} dx )$
associated with 
${\cal B}^2_\mu$ via a $\mu$-deformed Segal-Bargmann transform. 
(See \cite{SOGTMP} or \cite{MA} for more details.) 
In this space there is a naturally defined number operator
and its associated quadratic form. 
It seems reasonable to conjecture that there is a log-Sobolev inequality in this
space as well as a hypercontractivity result on the scale of Banach spaces
$L^p( {\bf R}, |x|^{2\mu} dx )$ for $ p > 1$.
Moreover, we conjecture that neither a reverse log-Sobolev inequality
nor a reverse hypercontractivity result holds in this context.

\vskip .4cm \noindent
\textbf{\Large Dedication}
\vskip .4cm \noindent
This work owes much to Marvin Rosenblum at a purely scientific level. 
(See \cite{RO}, a work chock full of interesting results.) 
But Marvin was also a wonderful teacher, from whom the second author learned 
a lot of analysis and operator theory, 
including his first ever introduction to the Segal-Bargmann space. 
The articles \cite{RKSBS}, \cite{RLSI}, \cite{SOME} and \cite{SOGTMP}
indicate just how important that introduction was for the second author. 
And our work in \cite{AS} owes much to \cite{RO}. 
The news of Marvin's death saddened us greatly. 
As a friend has remarked, ``He was one of the good guys.'' 
He certainly was. 
That alone is more than reason enough to dedicate this article to his memory.

\end{document}